\documentclass[lettersize,journal]{IEEEtran}
\usepackage{amsmath,amssymb,amsfonts}
\usepackage{algorithmic}
\usepackage{algorithm}
\usepackage{array}
\usepackage[caption=false,font=normalsize,labelfont=sf,textfont=sf]{subfig}
\usepackage{textcomp}
\usepackage{stfloats}
\usepackage[hyphens]{url}
\usepackage{makecell}
\usepackage{verbatim}
\usepackage{graphicx}
\usepackage{cite}
\usepackage[most]{tcolorbox}
\usepackage{layouts}
\usepackage{multirow}
\usepackage{siunitx}
\usepackage{tabularray}
\usepackage{float}
\usepackage{cases}
\usepackage[nice]{nicefrac}
\usepackage[autostyle]{csquotes} 

\usepackage{bm}
\usepackage{siunitx}
\usepackage{booktabs}

\usepackage{xargs}                      
\usepackage{fancyhdr}
\usepackage[colorinlistoftodos,prependcaption,textsize=tiny]{todonotes}

\newcommandx{\panqi}[2][1=]{\todo[linecolor=yellow,backgroundcolor=yellow!25,bordercolor=yellow,#1]{#2}}

\newcommandx{\topic}[2][1=]{\todo[linecolor=red,backgroundcolor=red!25,bordercolor=red,#1]{#2}}
\usepackage{threeparttable}
\usepackage{blindtext}
\usepackage{hyperref}
\hypersetup{hidelinks}
\hypersetup{
	colorlinks   = true, 
}
\usepackage[capitalize]{cleveref}
\usepackage{orcidlink}
\newcommand{\emailaddress}{panqi.jia@fau.de}
\usepackage{tikz}
\definecolor{myorange}{RGB}{181, 115, 91}
\definecolor{myblue}{RGB}{125,161,177}
\definecolor{mygreen}{RGB}{151, 183, 150}
\definecolor{mygray}{RGB}{156,156,156}

\usepackage{pifont}
\newcommand{\RomanNumeralCaps}[1]
    {\MakeUppercase{\romannumeral #1}}
\begin{document}
	
\title{Overview of Variable Rate Coding in JPEG AI}

\author{Panqi Jia\orcidlink{0009-0002-5480-0137}, \and Fabian Brand\orcidlink{0000-0002-2022-1033}, \and Dequan Yu,  \and Alexander Karabutov, \and Elena Alshina\orcidlink{0000-0001-7099-5371}, \and André Kaup \orcidlink{0000-0002-0929-5074}~\IEEEmembership{Fellow,~IEEE}
	\thanks{...}
	\thanks{Panqi Jia, Fabian Brand and André Kaup are with the Friedrich-Alexander University Erlangen-Nürnberg, Department of Electrical-Electronic-Communication Engineering, Chair of Multimedia Communications and Signal Processing, Erlangen 91058, Germany. (e-mail: \href{mailto:\emailaddress}{\emailaddress}).
 

		Panqi Jia, Fabian Brand, Dequan Yu,  Karabutov Alexander and Elena Alshina are with the Huawei Munich Research Center, Munich 80992, Germany.}
}


\IEEEpubid{0000--0000/00\$00.00~\copyright~2021 IEEE}

\maketitle
\thispagestyle{fancy}
\lhead{}
\lfoot{}
\cfoot{Copyright © 20xx IEEE. Personal use of this material is permitted. However, permission to use this material for any other purposes must be obtained from the IEEE by sending an email to pubs-permissions@ieee.org.}
\rfoot{}
\begin{abstract}
Empirical evidence has demonstrated that learning-based image compression can outperform classical compression frameworks. This has led to the ongoing standardization of learned-based image codecs, namely  Joint Photographic Experts Group (JPEG) AI. The objective of JPEG AI is to enhance compression efficiency and provide a software and hardware-friendly solution. Based on our research, JPEG AI represents the first standardization that can facilitate the implementation of a learned image codec on a mobile device.      
This article presents an overview of the variable rate coding functionality in JPEG AI, which includes three variable rate adaptations: a three-dimensional quality map, a fast bit rate matching algorithm, and a training strategy. The variable rate adaptations offer a continuous rate function up to 2.0 bpp, exhibiting a high level of performance, a flexible bit allocation between different color components, and a region of interest function for the specified use case. The evaluation of performance encompasses both objective and subjective results. With regard to the objective bit rate matching, the main profile with low complexity yielded a 13.1\% BD-rate gain over VVC intra, while the high profile with high complexity achieved a 19.2\% BD-rate gain over VVC intra. The BD-rate result is calculated as the mean of the seven perceptual metrics defined in the JPEG AI common test conditions. With respect to subjective results, the example of improving the quality of the region of interest is illustrated.

\end{abstract}

\begin{IEEEkeywords}
	Learned Image Compression, Variable Rate, Bit Rate Matching, Software and Hardware Friendly
\end{IEEEkeywords}

\section{Introduction}
\IEEEPARstart{t}{he} consumption of online media is increasing, leading to a growing need for high-quality content with lower bitrates. The increasing need for media has resulted in the swift advancement of traditional and neural network-based image codecs~\cite{ma2019image,birman2020overview}. In terms of rate distortion performance, the early NN-based codecs~\cite{balle2017end,balle2018variational,minnen2018joint,liu2019non,zhou2019multi,lee2018context,mentzer2018conditional,cheng2020learned,qian2020learning,9067005,koyuncu2021parallel,he2021checkerboard,minnen2020channel,qian2021entroformer,li2021deepcontextualvideocompression,Ladune2020OpticalFA,jia2022learningbased,Choi2019VariableRD,song2021variable,9522770,Cui2020GVAEAC,cui2021asymmetric,wang2023evcrealtimeneuralimage,10274142,VR_side} are comparable to traditional standards such as JPEG~\cite{wallace1992jpeg}, JPEG2000~\cite{skodras2001jpeg} and BPG~\cite{bellard2015bpg} (an intra coding of HEVC~\cite{sze2014high}). Furthermore, the recently proposed NN-based image codecs~\cite{he2022elic,guo2021causal,koyuncu2022contextformer,effcient_context} have already surpassed the intra coding of state-of-the-art video coding standards, such as VVC~\cite{ohm2018versatile}. 

In 2019, the JPEG standardization committee initiated a new project called JPEG AI~\cite{ascenso2021white}. In~\cite{ascenso2021white}, the scope of JPEG AI is described as follows: 
\blockquote{ The scope of JPEG AI is to create a learning-based image coding standard that provides a single-stream, compact compressed domain representation targeting both human visualization, with a significant improvement in compression efficiency over commonly used image coding standards at equivalent subjective quality, and effective performance for image processing and computer vision tasks, with the goal of supporting a royalty-free baseline.}
The development of the project is the result of the joint efforts of the International Electrotechnical Commission (IEC), the International Organization for Standardization (ISO), and the International Telecommunication Union (ITU). An overview of existing solutions~\cite{wg1n83058} showed that neural-based solutions are mature enough to compete with conventional image codecs. In light of this, the Call for Evidence~\cite{wg1n86018,wg1n89022} was established in 2020, followed by the Call for Proposals (CfP) in 2022~\cite{wg1n100095}. 
Ten teams participated in CfP and demonstrated superior results~\cite{wg1n100250} to those achieved by VVC~\cite{ohm2018versatile} in terms of the averaged metrics described in the Common Test and Training Conditions (CTTC)~\cite{wg1n100106}. 
A combination of the various proposals was used as the code base for the further development of the Verification Model (VM)~\cite{wg1n100279}. According to the standard's timeline, the issuance of the International Standard (IS) is projected for early 2025. During the collaboration phase, the decoder component of the verification model of the standard was ported to a mobile device~\cite{wg1n100658}, demonstrating the feasibility of decoding 4k images with high quality in real time.
\IEEEpubidadjcol

JPEG AI offers not only high compression efficiency but also flexible variable rate adaptation. The following techniques pertain to the variable rate adaptation in JPEG AI. The previously proposed conditional color separation (CCS) framework\cite{jia2022learningbased} is employed as the foundational framework within the JPEG AI standardization VM. The CCS framework is capable of generating two bitstreams for the different color components, thereby providing the ability to assign bits flexibly on different color components. 
Furthermore, a channel-wise quality map is employed to assist in the overall rate control. This map is based on the \textit{gain unit}, which is proposed in \cite{Cui2020GVAEAC}. The gain unit is a flexible matrix that can assist a neural network-based codec in reaching a continuous variable rate. Additionally, the efficient bit rate matching (BRM) algorithm enables JPEG AI to achieve a flexible target rate point. Moreover, supplementary spatial quality maps~\cite{jia2024bitdistributionstudyimplementation} are utilised within JPEG AI to facilitate variable rate functionality within the spatial domain. At last, a variable rate adaptive training strategy offers JPEG AI the potential to perform flexible rate coding.

In order to evaluate the BD-rate performance, seven quality metrics are employed, including MS-SSIM~\cite{1292216}, Visual Information Fidelity (VIF)~\cite{citVif}, feature similarity (FSIM)~\cite{5705575}, Normalized Laplacian Pyramid (NLPD)~\cite{nlpdcite}, Information Content Weighted Structural Similarity Measure (IW-SSIM)~\cite{5635337}, Video Multimethod Assessment Fusion (VMAF)~\cite{VMAFcite}, and PSNR-HVS-M~\cite{psnrHVS}. The main profile of JPEG AI has been demonstrated to achieve, on average across seven metrics, a 13.1\% BD-rate gain over the VVC intra mode. Furthermore, the high profile with high complexity has demonstrated the capacity to achieve a 19.2\% BD-rate gain over VVC intra on average across seven metrics. Moreover, the spatial quality map can enhance the subjective quality of the region of interest (ROI).

The following section presents an overview of related works in the field of learned image codecs and variable rate functions. Section \RomanNumeralCaps{3} outlines the variable rate adaptions in JPEG AI. Section \RomanNumeralCaps{4} describes the experiments and the experimental results, while Section \RomanNumeralCaps{5} presents the conclusions.

The following functions of JPEG AI are the focus of this article: 

\begin{description}
  \item[$\bullet$] A 3D quality map of the identical size to that of the latent tensor can be generated through the extension of a channel-wise quality map, a spatial quality map, or a combination of both.
  \item[$\bullet$] The channel-wise quality map is employed to facilitate an overall rate control, thereby enabling the model to achieve a continuous variable rate. By collaborating with an efficient bit rate alignment algorithm, the model is capable of precisely matching to a targeted rate point with optimal performance. 
  \item[$\bullet$] A spatial quality map is employed as a means of facilitating flexible spatial bit allocation. To reduce the number of bits used to signal the spatial quality map, only the residual of a spatial quality map with integer elements is transmitted. 
  \item[$\bullet$] A training strategy comprising distinct stages was introduced, with each stage encompassing a unique training parameter. This approach was designed with the objective of enhancing overall performance and facilitating variable rate adaptation.
\end{description}

\begin{figure}
    \centering
    \includegraphics[width=0.45\textwidth]{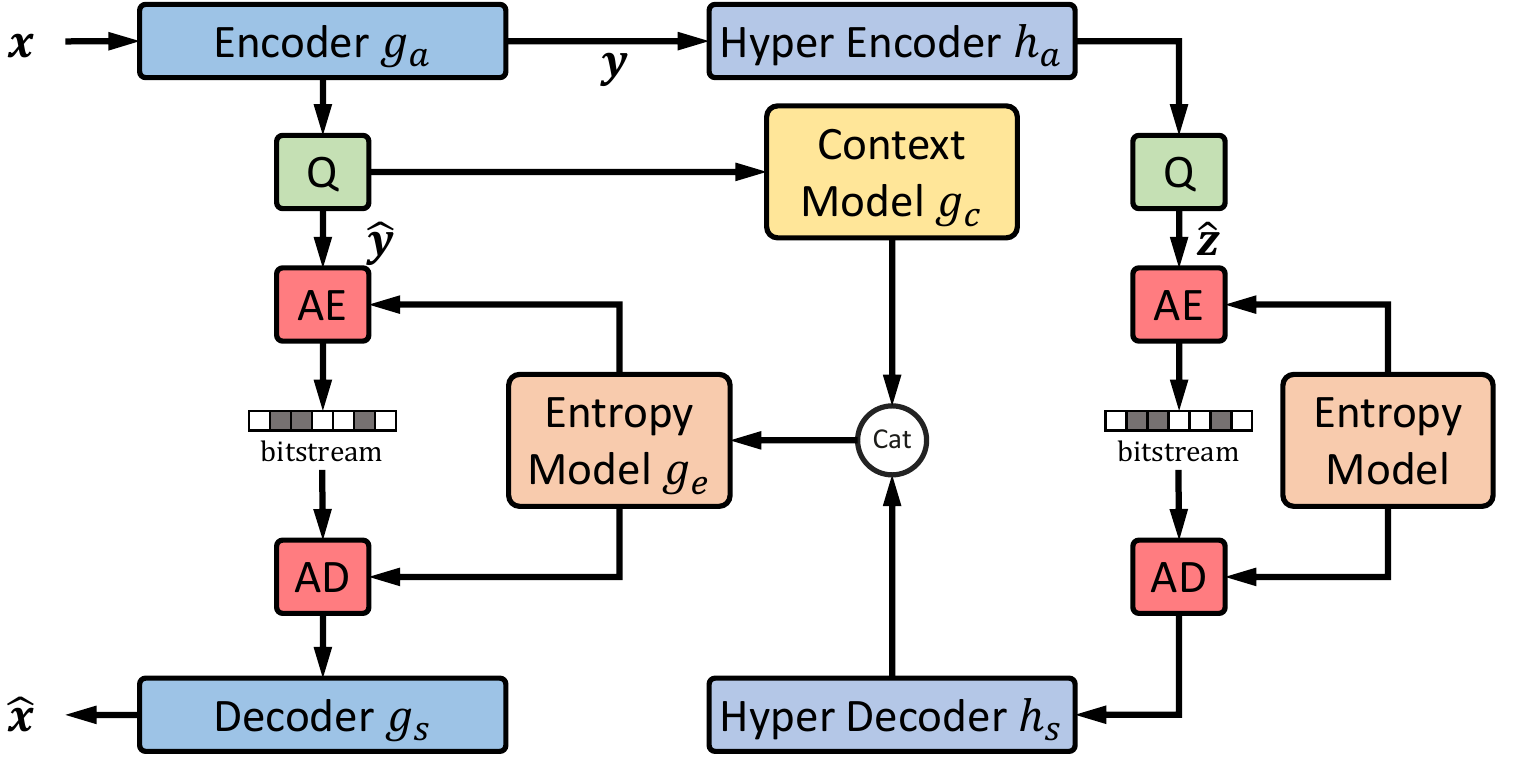}
    \caption{The classic autoencoder framework with hyperprior model, as proposed by Minnen et al.~\cite{minnen2018joint}, employs a quantization function, represented by $\rm{Q}$, and a lossless entropy codec, represented by $\rm{AE}$ and $\rm{AD}$. $\rm{Cat}$ indicates the concatenation.}
    \label{fig:autoencoder}
\end{figure}

\section{Related Work}
\subsection{Autoencoder in Learned Image Compression}
\label{sec_1}

Minnen et al. proposed a learned image compression codec in~\cite{minnen2018joint}. As illustrated in the Fig. \ref{fig:autoencoder}, their proposed codec has an autoencoder-like structure and consists of \textit{encoder, decoder, hyperprior, context model} and \textit{entropy model}. The encoder employs a non-linear analysis transform $g_a$ to transform the input image $\bm{x}$ into the latent tensor $\bm{y}$. This is followed by a quantization step $Q$, which quantizes the  latent tensor $\bm{y}$ to $\bm{\hat{y}}$. The encoding process is described by: 
\begin{equation}
     \bm{\hat{y}}= Q(g_a(\bm{x}; \bm{\phi_a})),
    \label{eq:encoder}
\end{equation}
where $\bm{\phi_a}$ are the parameters of the encoder.

Subsequently, the quantized $\bm{\hat{y}}$ will be encoded into a bitstream using a lossless entropy coder. The decoder receives $\bm{\hat{y}}$ from the bitstream and employs a non-linear synthesis transform $g_s$ to reconstruct the output image $\bm{\hat{x}}$ from the $\bm{\hat{y}}$. The decoder can be formulated as
\begin{equation}
    \bm{\hat{x}}= g_s(\bm{\hat{y}}; \bm{\phi_s})),
    \label{eq:decoder}
\end{equation}
where $\bm{\phi_s}$ are the decoder parameters.

In order to transmit the bits as efficiently as possible, the entropy model must learn the probability distribution of $\bm{\hat{y}}$. The forward and backward adaption proposed in~\cite{minnen2020channel} can assist the entropy model in this endeavour. A hyperprior network is employed for forward adaption. The hyperprior is another autoencoder-like network comprising an analysis transform $h_a$ and a synthesis transform $h_s$, which generates a hyper latent tensor $\bm{z}$. This is then quantized to $\bm{\hat{z}}$ as illustrated by
\begin{equation}
		\bm{\hat{z}}= Q(h_a(\bm{\hat{y}}; \bm{\theta_a})),
        \label{eq:hyper}
\end{equation}
where $\theta_a$ are the parameters of the hyper encoder.

 The backward adaption comprises an autoregressive context model $g_c$, which is capable of estimating the entropy of the latent tensor $\bm{\hat{y_i}}$ using the previously coded elements $\bm{\hat{y}}_{<i}$. Furthermore, the entropy model $g_e$ utilises the parameters generated by the context model $g_c$ and hyperprior to construct the conditional distribution $p_{\bm{\hat{y}}}(\bm{\hat{y}}|\bm{\hat{z}})$. This process is described by
\begin{equation}
  p_{\bm{\hat{y}_{i}}}(\bm{\hat{y}_{i}}|\bm{\hat{z}})\leftarrow g_{e}( g_{c}(\bm{\hat{y}}_{<i}; \bm{\theta_{c}}), h_s(\bm{\hat{z}}; \bm{\theta_s}); \bm{\psi_{e}}),
		\label{eq:all}
\end{equation}
where $\bm{\theta_{c}}$ represents the context model parameters,  $\bm{\theta_{s}}$ indicates the hyper decoder parameters, and $\bm{\psi_{e}}$ is the parameters of the entropy model $g_e$.

The whole codec can be trained end-to-end, and the loss function of the training process is: 
\begin{subequations}
	\begin{align}
		\mathcal{L}(\bm{\phi}, \bm{\theta}, \bm{\psi}) &= \mathbf{R}(\bm{\hat{y}}) + \mathbf{R}(\bm{\hat{z}}) + \beta_\mathrm{train} \cdot \mathbf{D}(\bm{x}, \bm{\hat{x}}),\label{eq_2}\\
		&= \mathbb{E}[-\log_2(p_{\bm{\hat{y}}}(\bm{\hat{y}}|\bm{\hat{z}}))] + \mathbb{E}[-\log_2(p_{\bm{\hat{z}}}(\bm{\hat{z}}|\bm{\psi}))]\notag\\
		&\quad + \beta_\mathrm{train} \cdot \mathbf{D}(\bm{x}, \bm{\hat{x}}).
	\end{align}
\end{subequations}
The optimization parameters $\phi$, $\theta$ and $\psi$ correspond to the transforms under consideration. The Lagrange multiplier $\beta_\mathrm{train}$ serves to regulate the trade-off between distortion $\mathbf{D}(\cdot)$ and rate $\mathbf{R}(\cdot)$.

\subsection{Conditional Color Separation (CCS) Framework}
The use of auxiliary information in the autoencoder can help to improve the coding efficiency. This concept has been successfully applied in both video and image compression domains, as proposed in~\cite{li2021deepcontextualvideocompression,Ladune2020OpticalFA,9244548,jia2022learningbased}.
\begin{figure}
    \centering
    \includegraphics[width=0.45\textwidth]{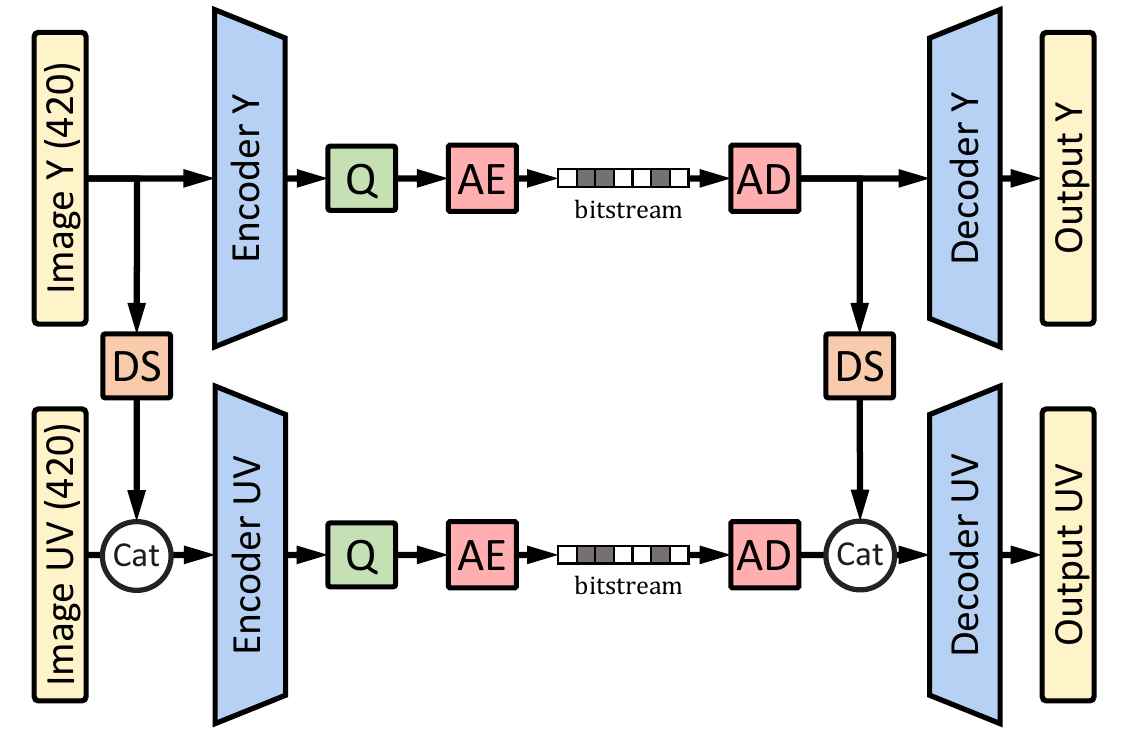}
    \caption{An example illustrates the processing of YUV420 format input using the CCS framework. The Y and UV components are separated into two groups, with Y considered the primary color component and UV the secondary component. $\rm{DS}$ indicates the down-sampling process, which is necessary because Y has a larger spatial size than UV.}
    \label{fig:CCS}
\end{figure}
The results of the studies conducted by~\cite{Ladune2020OpticalFA} and~\cite{9244548} demonstrate that conditioning an image autoencoder yields superior compression ratios and enhanced reconstruction quality. In light of the works~\cite{Ladune2020OpticalFA} and~\cite{9244548}, we proposed a conditional autoencoder framework for image compression, which we term CCS. In our CCS framework, the primary and auxiliary information are derived from distinct color channels. This approach involves the use of a primary color component to facilitate the processing of non-primary color components. 

JPEG AI has adopted CCS as its foundational framework. Fig. \ref{fig:CCS} shows an example of CCS applied to the JPEG AI. In this example, the input data is transformed from the RGB to the YUV420 format. Note that JPEG AI can support different internal YUV formats. Here, we just show an example of using YUV420. The YUV format has three components: Y contains the luminance, while U and V contain the chrominance information. In the YUV420 format, the U and V components are sub-sampled by a factor of two in each spatial direction. Given that the Y component has four times more samples, it is selected to be the primary component. 

Prior to processing, the non-primary components U and V are concatenated, and the primary component Y is down-sampled to match the height and width of the concatenated UV components. Each group of components is processed by a dedicated encoder, decoder, hyperprior, autoregressive context model, arithmetic encoder, and arithmetic decoder. The processing of Y occurs in parallel with the processing of UV, resulting in the generation of two distinct bitstreams. The UV components are coded with the Y component's auxiliary information. The Y component and its latent representation will be concatenated with the UV components before the encoder and the decoder. The Y component will be downsampled before concatenation, when Y and UV have different sizes.


The CCS framework in JPEG AI enables the generation of dedicated bitstreams for the Y and UV components, thereby facilitating a more flexible variable rate adaptation in accordance with the varying color components.

\subsection{Gain Unit Based Variable Rate Adaption}
The gain unit~\cite{Cui2020GVAEAC,cui2021asymmetric} proposed by Cui et al. has the potential to assist the learned image codec in achieving a continuous rate.
\begin{figure}
    \centering
    \includegraphics[width=0.48\textwidth]{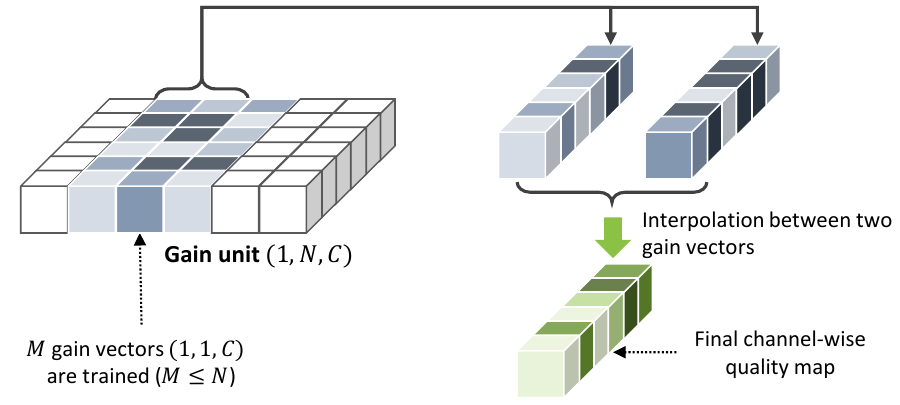}
    \caption{The illustration of the gain unit is as follows: $C$ represents the channel number in the latent space, $N$ denotes the total number of gain vectors in the gain unit, $M$ signifies the number of updated gain vectors for one model during the training process, and $ft$ represents the interpolation parameter between two gain vectors.}
    \label{fig:Gain_unit}
\end{figure}
Fig. \ref{fig:Gain_unit} depicts the gain unit~\cite{Cui2020GVAEAC}, which is a trainable matrix comprising multiple gain vectors. Each gain vector provides a channel-wise quality map, offering a range of quantization steps for each channel. During the training phase, each model with its dedicated set of gain vectors will be trained together. Each model is capable of achieving multiple discrete rate points by utilizing its dedicated set of gain vectors. Moreover, the application of exponential interpolation between two neighboring gain vectors enables the generation of a channel-wise quality map that can provide the rate between two discrete rate points. The process of exponential interpolation between two gain vectors is defined by the following equation:
\begin{equation}
    \bm{m}_f = {\bm{m}_{n}}^{ft} \cdot {\bm{m}_{n+1}}^{1-ft},  
\end{equation} 
where $ n \in$  [0, 1, $\cdot \cdot \cdot$ , N-1] , $N$ is the total number of gain vectors within the gain unit. $m_f$ denotes the final channel-wise quality map, which is obtained through exponential interpolation between the gain vectors $m_n$ and $m_{n+1}$. Additionally, ft represents the interpolation parameter. 

In JPEG AI, a simplified gain unit is employed to facilitate the implementation of a variable rate in a manner that is compatible with both software and hardware. This approach will be detailed in the subsequent section.

\begin{figure}
    \centering
    \setkeys{Gin}{width=0.99\linewidth} 

\subfloat[\label{fig:ChannelQ}]{\includegraphics[width=0.99\linewidth]{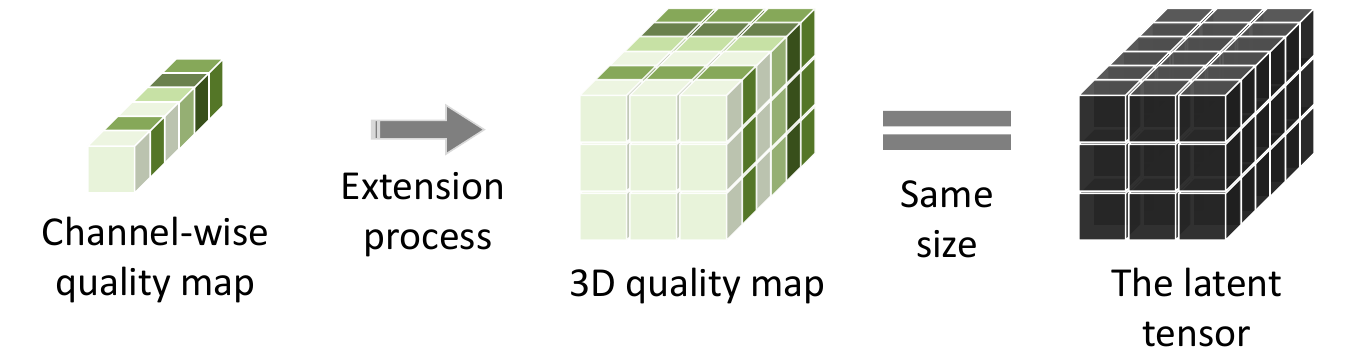} }\hfil
\subfloat[\label{fig:SpatialQ}]{\includegraphics[width=0.99\linewidth]{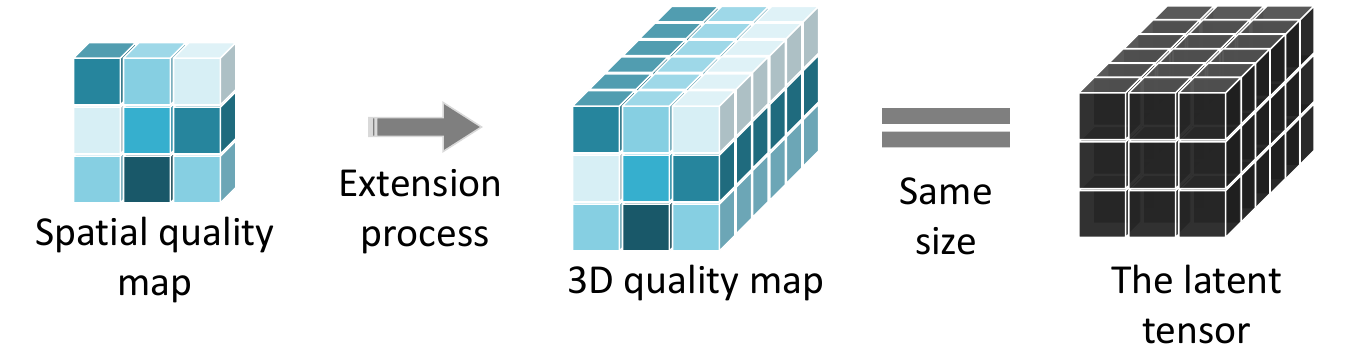} }\hfil
\subfloat[\label{fig:JointQ}]{\includegraphics[width=0.99\linewidth]{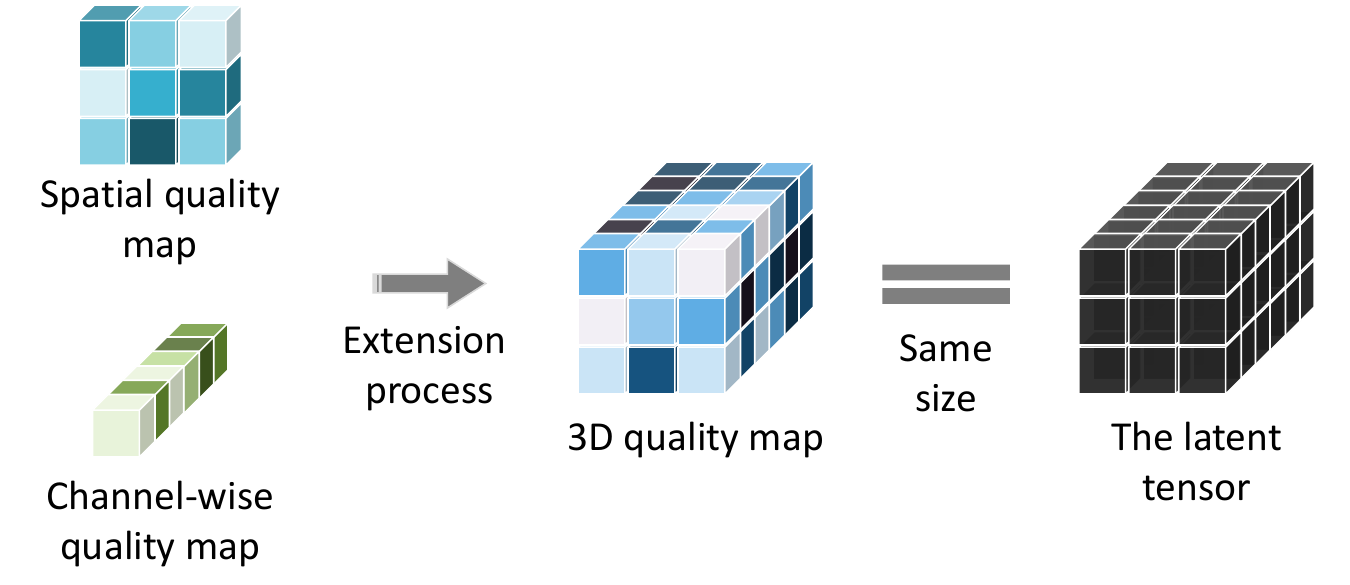} }\hfil

\caption{Fig. \ref{fig:ChannelQ} demonstrates the extension of the channel-wise quality map to a 3D quality map. Fig. \ref{fig:SpatialQ} demonstrates the extension of the spatial quality map to a 3D quality map, and Fig. \ref{fig:JointQ} shows the utilisation of a combined spatial and channel-wise quality map, extending to a 3D quality map.}
\label{fig:extension}
\end{figure}

\section{Variable Rate functionality}
This section will present an overview of the variable rate adaptation functionality available in JPEG AI. The first step will be to explain the three-dimensional (3D) quality map, which offers a flexible variable rate in JPEG AI. The 3D quality map can be generated by extending either the channel-wise or the spatial quality map. An alternative approach is to extend two kinds of maps jointly. Subsequently, the channel-wise quality map and the spatial quality map will be presented in separate sections. The channel-wise quality map is employed to regulate the overall bit rate. The spatial quality map enables the distribution of bits to be modified in accordance with the ROI. Subsequently, the proposed BRM algorithm in JPEG AI will be introduced. This algorithm utilizes the channel-wise quality map to achieve a precise rate point. Finally, the training strategy will be presented to facilitate the variable rate.

\subsection{Three-Dimensional Quality Map}
\label{sec:3dmap}

JPEG AI first converts the input RGB data into different YUV formats. Then the internal process will encode the YUV format because the CCS framework allows to set the Y component as the primary component. The input RGB image has dimensions $(H, W, 3)$, when converted to YUV, the luminance has dimension $(H, W, 1)$, and the chrominance components can have different spatial dimensions, such as $(\nicefrac{H}{2},\nicefrac{W}{2}, 2)$ or $(H, W, 2)$, and so on. Note that for different YUV formats, the chrominance encoder pre-processes the input by using an unshuffle layer. The input of the chrominance encoder contains the auxiliary information of Y. This pre-processing can make the input of the chrominance layer always have dimension $(\nicefrac{H}{2},\nicefrac{W}{2}, C_\mathrm{UV}^i)$, where $C_\mathrm{UV}^i$ denotes the channels after the unshuffle layer, and it can be different for the different YUV formats. After the encoding process, the latent representation of luminance has a size of $(\nicefrac{H}{16}, \nicefrac{W}{16}, C_\mathrm{Y})$, while the latent representation of chrominance has a size of $(\nicefrac{H}{16}, \nicefrac{W}{16}, C_\mathrm{UV})$. $H$ indicates the height and $W$ indicates the width of the image, $C_\mathrm{Y}$ represents the luminance channel number in the latent space and $C_\mathrm{UV}$ represents the chrominance channel number in the latent space.

The variable rate functionality in JPEG AI uses different scaling for each element of the latent tensor, which requires the use of two three-dimensional (3D) quality maps with dimensions $(\nicefrac{H}{16}, \nicefrac{W}{16}, C_\mathrm{Y})$ and $(\nicefrac{H}{16}, \nicefrac{W}{16}, C_\mathrm{UV})$. In addition, signaling 3D quality maps with the same size as the latent tensor resulted in an excess of bits. To circumvent this problem, the 3D quality map is generated by extending either a channel-wise quality map or a spatial quality map, and only the control parameters of two kinds of quality maps are signaled. The control parameter of the channel-wise quality map is $\Delta_{\beta}$, a 12-bit integer value which is written in the picture header. Given that different components have their own dedicated bitstream, it is able to allocate bits between components. As a result, there will be two control parameters for luminance and chrominance, and they will usually be set to the same value. The details of $\Delta_{\beta}$ are presented in \ref{sec:Channel-wise}. 
The control parameter of the spatial quality map is Q with dimension $(\nicefrac{H}{16}, \nicefrac{W}{16}, 1)$, which is  written in its own substream when required. Since the spatial dimension for luminance and chrominance in the latent space is identical, only one unique spatial map will be used. Details on Q are introduced in \ref{sec:spatial}.

The extension processes of different types of quality maps are shown in Fig. \ref{fig:extension}. Fig. \ref{fig:ChannelQ} illustrates the process of extending the channel-wise quality map to a 3D quality map. A channel-wise quality map has a size of (1,1,C), and by extending the channel-wise quality map in the spatial domain, a 3D quality map can be generated. Fig. \ref{fig:SpatialQ} illustrates the process of extending the spatial quality map to a 3D quality map. This is achieved by duplicating the value in the channel domain, which generates the 3D quality map.

In the majority of use cases, the two types of quality maps are employed in isolation. The channel-wise quality map is primarily concerned with the overall rate control, whereas the spatial quality map is utilised for the ROI. The specifics of the two quality map types are outlined in the following subsections. However, the two quality map types can also be employed in conjunction. In this particular instance, the spatial quality map must be multiplied by the disparate values in the channel domain to generate the 3D quality map. The process is illustrated in Fig. \ref{fig:JointQ}. 

The process of using a 3D quality map in the latent space to provide a variable rate is described by the equation:
\begin{equation}
    r^\prime_\mathrm{Y/UV}[c,i,j] = m_\mathrm{Y/UV}[c,i,j] \cdot r_\mathrm{Y/UV}[c,i,j], 
\end{equation}
where $0\leq i \leq \nicefrac{H}{16}, 0\leq j \leq \nicefrac{W}{16}, $ and $0\leq c \leq C_\mathrm{Y/UV}$. The subscript $_\mathrm{Y/UV}$ denotes the processing components, which can be either Y or UV.
The residual latent tensor, denoted by $r^\prime_\mathrm{Y/UV}$, is written into the bitstream and is computed by multiplying the original residual tensor $r_\mathrm{Y/UV}$ by the 3D quality map $m_\mathrm{Y/UV}$. The original tensor $r_\mathrm{Y/UV}$ is the result of subtracting the mean value from the latent tensor as : \begin{equation}
 r_\mathrm{Y/UV} = y_\mathrm{Y/UV} - \mu_\mathrm{Y/UV}.
 \end{equation}
The latent tensor $y_\mathrm{Y/UV}$ is generated by the encoder, and its mean value $\mu_\mathrm{Y/UV}$, is generated by the latent domain prediction. The latent domain prediction is a decoupled framework that was proposed by Zhang et al. in~\cite{10247017} and adopted in JPEG AI.

\subsubsection{Channel-Wise Quality Map in JPEG AI }
\label{sec:Channel-wise}
In JPEG AI, to cover a large bit rate range, there are four models trained for a specific Lagrange multiplier $\beta_\mathrm{train}$ determining rate-distortion trade-off as described in (\ref{eq_2}). A higher value of $\beta_\mathrm{train}$ yields a higher quality reconstructed image, requiring more bits to code. Each model in JPEG AI employs a dedicated gain unit to facilitate continuous variable rate coding. 
\begin{figure}
    \centering
    \includegraphics[width=0.48\textwidth]{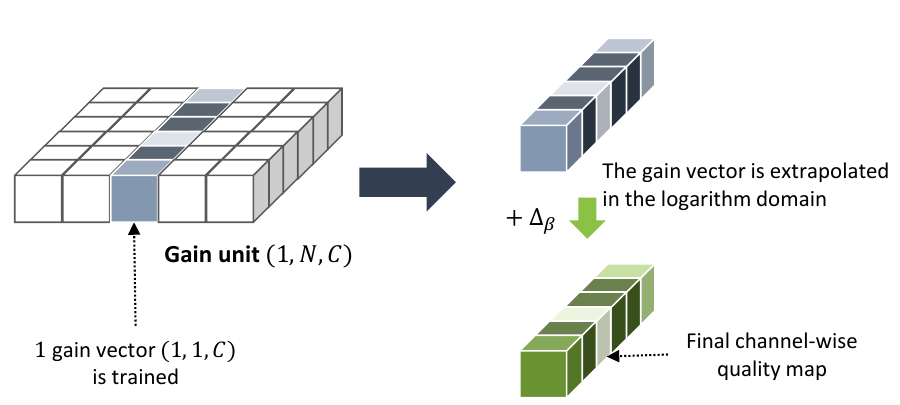}
    \caption{The illustration of the gain unit is as follows: $C$ represents the channel number in the latent space, $N$ denotes the total number of gain vectors in the gain unit, and $\Delta_{\beta}$ represents the extrapolation parameter for the single gain vector in the logarithm domain.}
    \label{fig:Gain_unit_simp}
\end{figure}
As illustrated in Fig. \ref{fig:Gain_unit_simp}, in contrast to the aforementioned gain unit~\cite{Cui2020GVAEAC}, the gain unit in JPEG AI comprises a single gain vector, which is used to generate the channel-wise quality map through extrapolation. In a practical application, the generation of a channel-wise quality map through exponential interpolation between two gain vectors requires a high computational complexity due to the necessity of searching for two gain vectors. In such a case, the utilization of a single gain vector circumvents the necessity for this search process, thereby facilitating an efficient implementation of gain unit JPEG AI.

In the practical application of learned image codecs on diverse devices, a significant challenge emerges in~\cite{10448359} and~\cite{10018040}, namely device interoperability. This arises from the use of floating-point calculations, where different devices exhibit disparate designs for presenting floating-point values. Consequently, some devices may be unable to decode the data accurately, due to discrepancies in the floating-point representations across different devices. In order to address the issue of device interoperability, the quantisation to entropy networks proposed in~\cite{10448359} and~\cite{10018040} have been implemented in JPEG AI. Consequently, all entropy-related calculations are now performed in integer precision.

To further simplify the hardware implementation and achieve greater speed, in JPEG AI all entropy-related calculations were placed into the logarithm domain. This was done with the objective of converting multiplication or exponentiation to addition. Following these modifications, the exponential extrapolation of the gain vector has been replaced with an addition operation, thereby facilitating the generation of the channel-wise quality map from the gain vector. During encoding, a $\beta_\mathrm{test}$ is required for the purpose of providing a variable rate. The $\beta_\mathrm{test}$ is initially used to calculate the displacement, $\delta_{\beta}$, via the following equation: 
\begin{equation}
\delta_{\beta} = \frac{\beta_\mathrm{test}}{\beta_\mathrm{train}}.
\end{equation}
 
Subsequently, the $\delta_{\beta}$ will be transformed into the $\Delta_{\beta}$ in the logarithmic domain through the application of the following equation:
\begin{equation}
    \Delta_{\beta} = \lfloor \ln{\delta_{\beta}} \cdot \frac{P_{\beta}}{S_{\sigma}} \rfloor,
\end{equation}
where $S_{\sigma}$ is the quantization step for the entropy model and $P_{\beta}$ is the precision length for the decimal of $\Delta_{\beta}$. In JPEG AI, the value of $S_{\sigma}$ was set to 0.2. The value of $P_{\beta}$ is equal to $2^7$, as a seven-bit integer of $\Delta_{\beta}$ is employed to describe the decimal of $\delta_{\beta}$. By multiplying $\nicefrac{P_{\beta}}{S_{\sigma}}$, the integer $\Delta_{\beta}$ can be extended to a sufficient length to describe the floating point $\delta_{\beta}$. At last, $\Delta_{\beta}$ will be added to the gain vector in order to obtain the channel-wise quality map. The aforementioned addition in the logarithm domain is equal to the multiplication in the normal domain; this addition calculation represents the extrapolation of the gain vector. 

The addition of different $\Delta_{\beta}$ to the gain vector allows for the generation of a channel-wise quality map for the purpose of overall rate control. The utilisation of a negative $\Delta_{\beta}$ results in a reduction in rate, whereas the application of a positive $\Delta_{\beta}$ facilitates an increase in rate. In conclusion, the channel-wise quality map enables the achievement of different rate points through the signaling of a single parameter $\Delta_{\beta}$, which requires only addition calculations.

\subsubsection{Spatial Quality Map in JPEG AI }
\label{sec:spatial}
JPEG AI employs a spatial bit distribution methodology that utilises a spatial quality map. This map assigns different quantization steps to the various areas of the image, thereby enabling the allocation of more bits to areas assigned a lower quantization step, which in turn facilitates a superior reconstructed quality. Conversely, areas assigned a higher quantization step receive fewer bits, resulting in a lower reconstructed quality. The spatial quality map within the codec is predefined in order to guarantee accurate decoder processing. It thus follows that the quality map must be incorporated into the bit stream in order to ensure correct decoding. Consequently, the spatial quality map is also subjected to an entropy encoding process.

JPEG AI compresses the input image $x$ with dimensions $(H, W, 3)$ into a latent tensor $y$. This latent tensor contains two distinct sets of components: the primary components, with dimensions $(\nicefrac{H}{16}, \nicefrac{W}{16}, C_\mathrm{Y})$, and the secondary components, with dimensions $(\nicefrac{H}{16}, \nicefrac{W}{16}, C_\mathrm{UV})$. As stated in Section \ref{sec:3dmap}, it is noteworthy that the spatial dimensions of the primary and secondary components are identical in the latent space. Consequently, each element in the latent tensor represents a $16\times16$ block of the input image. To facilitate spatial bit adaptation, JPEG AI incorporates a spatial quality map $Q$ with dimensions $(\nicefrac{H}{16}, \nicefrac{W}{16},1)$. By multiplying the latent tensor with $Q$, the original image can be allocated a flexible block-wise bit distribution.

\begin{table*}[t]
	\centering
	\footnotesize
        \caption{Quantization index and its corresponding quantization scale }
	\label{tab2}
	\begin{tabular}{c|ccccccccccccccccc}
		\toprule
		\textbf{Q index}&\textbf{-8}&\textbf{-7}&\textbf{-6}&\textbf{-5}&\textbf{-4}&\textbf{-3}&\textbf{-2}&\textbf{-1}&\textbf{0}&\textbf{1}&\textbf{2}&\textbf{3}&\textbf{4}&\textbf{5}&\textbf{6}&\textbf{7}&\textbf{8}\\
		\midrule
		\makecell[t]{\textbf{Q scale}}
		
		&\makecell[t]{0.25}
		
		&\makecell[t]{0.3125}
		
		&\makecell[t]{0.375}
		
		&\makecell[t]{0.4375}
		
		&\makecell[t]{0.5}
		
		&\makecell[t]{0.625}
		
		&\makecell[t]{0.75}
		
		&\makecell[t]{0.875}
		
		&\makecell[t]{1}
		
		&\makecell[t]{1.25}
		&\makecell[t]{1.4375}
		&\makecell[t]{1.6875}
		&\makecell[t]{2}
		&\makecell[t]{2.4375}
		&\makecell[t]{2.875}
		&\makecell[t]{3.375}
		&\makecell[t]{4}
		\\
		\bottomrule
  
	\end{tabular}
	
\end{table*}

Given that the spatial quality map must be incorporated into the bit stream, it is imperative to shorten the bit depth of each element within the spatial quality map. Consequently, the spatial quality map employs integer values. These integer values are referred to as quantization indexes, which indicate the floating-point quantization scale. The encoder and decoder are able to determine the corresponding quantization step by referencing the quantization index in the spatial quality map through a look-up table. Table. \ref{tab2} presents the 17 quantization index values, accompanied by their respective quantization scale. The latent tensor will multiply the quantization scale derived from the spatial map, and then proceed with the quantization process. Consequently, a larger quantization scale will result in a lower quantization loss and a higher reconstructed quality with more bits.

Since the spatial quality map has to be known at the decoder, it needs to be included it in the bitstream. JPEG AI use a linear predictor for that purpose and transmit the residuals. For each element $q[i,j]$ in the spatial map, the predicted value $q_p[i,j]$ is calculated by the equation \eqref{eq_pred_q}:

\begin{equation}
\label{eq_pred_q}
  q_{p}[i,j] =\begin{cases}
    (q[i,j-1] + q[i-1,j])/2,& \text{if } i > 0, j > 0 \\ 
    q[i-1,j],              & \text{if } i > 0, j = 0 \\
    q[i,j-1],              & \text{if } i = 0, j > 0 \\
    0,                      & \text{if } i = 0, j = 0
  \end{cases}
\end{equation}
where $i\in[0,\nicefrac{H}{16}]$ and $j\in[0,\nicefrac{W}{16}]$. The residual value $\delta_{q}$ is calculated by :
\begin{equation}
 \delta_{q}[i,j] = q[i,j] - q_{p}[i,j].    
\end{equation}
In conclusion, the spatial quality map can facilitate flexible bit allocation in the spatial domain. By signaling the residual of a spatial quality map with integer elements, a minimal amount of bits is required for transmission.

\subsection{Bit Rate Matching in JPEG AI }
\label{Sec_BRM}
In the JPEG AI requirements, the BRM condition is deemed to be satisfied when the generated rate differs from the target rate by less than 10\%. Given that four models are trained in JPEG AI, the objective of BRM is to identify the model and the corresponding $\Delta_{\beta}$ values that align with the target rate. The BRM algorithm is designed with the variable rate feature in JPEG AI as its foundation. This section will present an overview of the variable rate feature and the BRM algorithm. Note that the BRM algorithm described in this section is a proof of concept. The BRM algorithm is part of the JPEG AI reference software and is not normative. Implementers are free to develop algorithms that are more appropriate for their applications.

\subsubsection{Variable Rate Feature in JPEG AI}
\label{sec:vr_feature}
\begin{figure}
    \centering
    \includegraphics[width=0.48\textwidth]{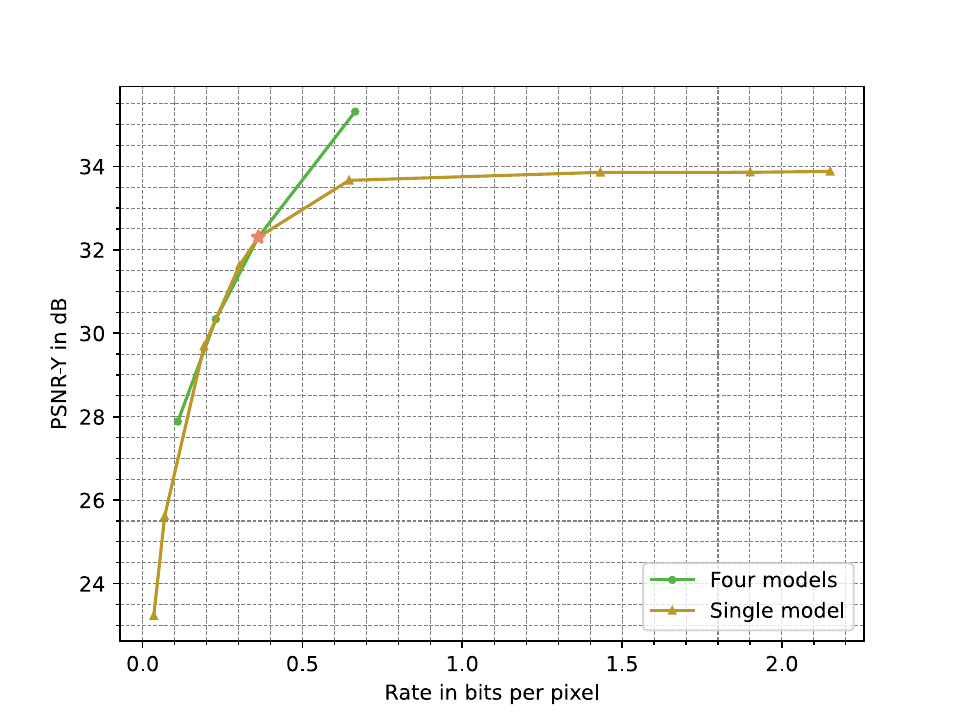}
    \caption{The illustration demonstrates the use of four models to generate an optimal variable rate curve and the use of a single model to cover a large bit rate range. }
    \label{fig:BDL}
\end{figure}

In JPEG AI, to cover a large bit rate range, there are four models trained for a specific Lagrange multiplier $\beta_\mathrm{train}$ determining rate-distortion trade-off as described in (\ref{eq_2}). A higher value of $\beta_\mathrm{train}$ yields a higher quality reconstructed image, requiring more bits to code. 
The rationale behind utilising four models is explained in Fig. \ref{fig:BDL}, which depicts two distinct lines. The green line illustrates the process of employing four models to construct an optimal variable rate envelope. The curve represents four points, each denoting a model that has been trained with a distinct $\beta_\mathrm{train}$ and encoding with the $\beta_\mathrm{test} = \beta_\mathrm{train}$. These points are referred to as anchor points.

As illustrated in Fig. \ref{fig:BDL}, the generation of multiple bit rate points is demonstrated by the yellow curve, which employs a single model. The yellow curve was generated using the single model indicated by a pink star on the optimal variable rate envelope. The pink star represents the anchor point of the third model from the left on the optimal variable rate envelope. From this figure, we can observe the following characteristics:    
\begin{description}
  \item[$\bullet$]It is feasible to utilise a single model to encompass a broad bit rate range.
  \item[$\bullet$]The single model exhibits optimal performance when the bit rate range is close to the anchor point.
  \item[$\bullet$]If the single model provides a bit rate that is either too low or too high in comparison to the anchor point, the resulting performance will not meet the desired specifications.
  \item[$\bullet$]When providing a higher bit rate than the anchor point from the single model, the performance decline is more rapid than when providing a lower bit rate.
  \item[$\bullet$]The method of providing a variable rate for a single model is based on the alteration of quantization steps. In the region of the higher bit rate, the BD-rate performance is already saturated when the rate is increased. This phenomenon can be explained by the fact that autoencoders are lossy systems even in the absence of quantization. Increasing the quantization step size will not result in improved model performance after a certain point.
\end{description}

In consideration of the aforementioned listed features from Fig. \ref{fig:BDL}, it can be observed that the use of a single model has the potential to cover a long bit rate range. However, this approach may also result in a significant variable rate loss. To address this challenge, JPEG AI employs a four-model approach, with each model designed to provide a specific bit rate range. This is achieved by limiting the value of the parameter $\Delta_{\beta}$, introduced in \ref{sec:Channel-wise}, within the interval $[-1069, 702]$. The lower and upper bounds of this interval are determined based on empirical observations. It should be noted that the lower bound has a greater absolute value than the upper bound. This is due to the fact that the conclusion drawn from Fig. \ref{fig:BDL}, which states that "When providing a higher bit rate than the anchor point from the single model, the performance decline is more rapid than when providing a lower bit rate," is a contributing factor.  

When the codec is required to achieve a specified target rate, a strategic approach is necessary to select an appropriate model and provide the corresponding $\Delta_{\beta}$ value. The following paragraph will introduce the BRM algorithm.

\subsubsection{Bit Rate Matching Algorithm in JPEG AI}
\label{sec:BRM_alg}
\begin{figure}
    \centering
    \includegraphics[width=0.48\textwidth]{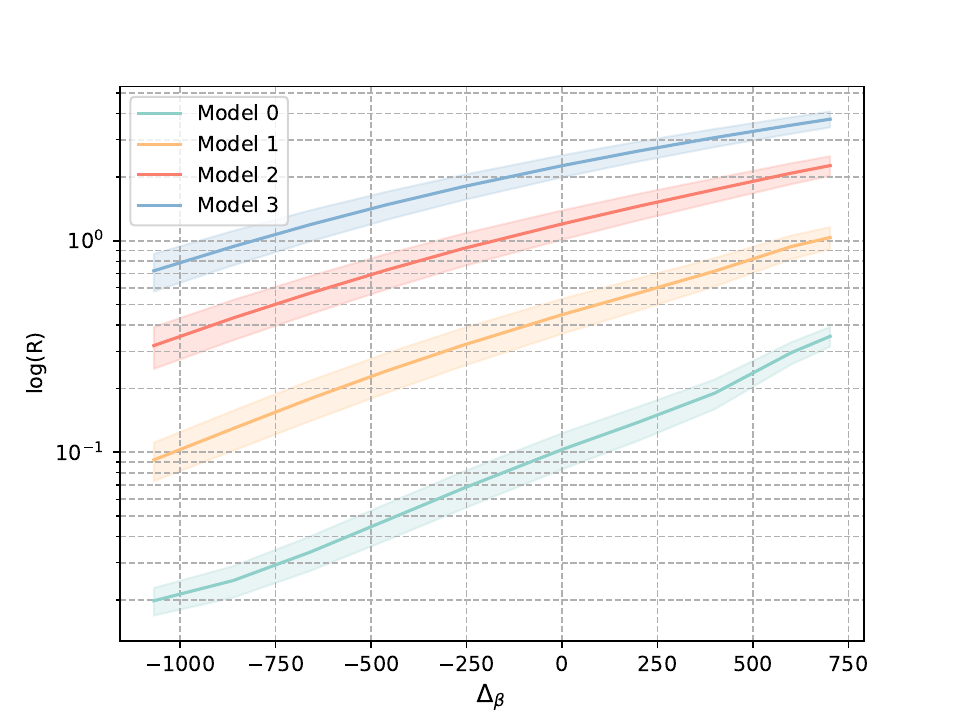}
    \caption{The illustration of applying distinct $\Delta_{\beta}$ values to the same model, using a channel-wise quality map, shows that varying $\Delta_{\beta}$ can facilitate the model's ability to provide disparate rate $R$. The relationship between ${\rm log}(R)$ and $\Delta_{\beta}$ is approximately linear. Four models of the main profile are shown.}
    \label{fig:LinearFunc}
\end{figure}

In JPEG AI, the BRM process is described in the previous proposed work~\cite{faucris.325748200}, it is comprised of three distinct parts: model selection, $\Delta_{\beta}$ searching, and $\Delta_{\beta}$ validation. In the model selection stage, BRM must locate the particular model that offers the target rate. In the $\Delta_{\beta}$ searching phase, the ultimate $\Delta_{\beta}$ is determined for the selected model. In the $\Delta_{\beta}$ validation phase, the codec computes the rate and loss for the $\Delta_{\beta}$. This subsection provides a detailed account of each BRM step.

In the context of model selection, the model exhibiting the closest relative bit distance $D_r$ to the target rate will be selected as the candidate. The following BRM steps will be applied exclusively to the candidate in order to achieve the target rate. The relative bit distance to target rate is calculated by:
\begin{equation}
    D_r = \frac{{\rm abs}(R_{d} - R_{t})}{R_{d}},
\end{equation}
where $R_{d}$ is used to denote the default rate of each model, which is generated using  $\beta_\mathrm{test} = \beta_\mathrm{train}$. By employing this approach, the model is biased to use a model with a higher default rate to generate a lower rate, which matches what we observed in Fig. \ref{fig:BDL}. It is necessary for all four models to calculate the $D_r$ to the target rate, and the model with the closest $D_r$ will be chosen as the candidate.

In order to find an appropriate $\Delta_{\beta}$ to achieve the target rate for the selected candidate, JPEG AI uses linear interpolation for the $\Delta_{\beta}$ search. As illustrated in Fig. \ref{fig:LinearFunc}, our observations indicated that the relationship between $\Delta_{\beta}$ and ${\rm log}(R)$ is approximately linear, which allowed us to hypothesize that:
\begin{equation}
    {\rm log}(R) \approx a \cdot \Delta_{\beta} + b  ,
\end{equation}
where $a$ and $b$ represent constant values, while $\Delta_{\beta}$ is the signaled quality parameter in logarithm domain that is mentioned in \ref{sec:Channel-wise}.

\begin{figure}
    \centering
    \includegraphics[width=0.48\textwidth]{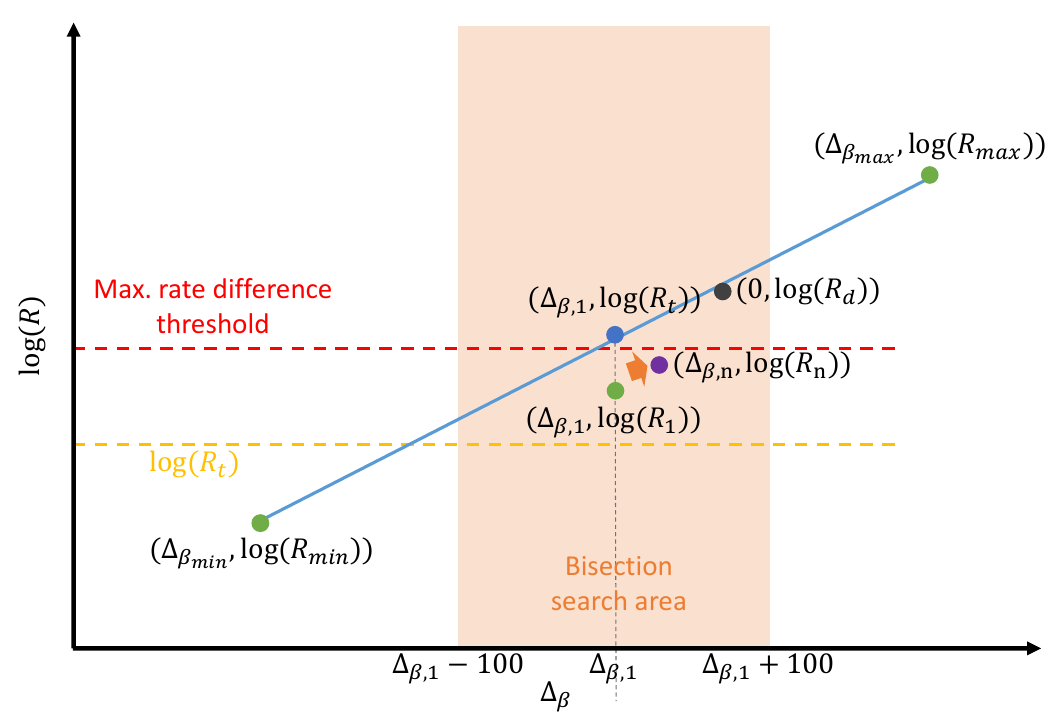}
    \caption{The illustration of the $\Delta_{\beta}$ search employs a linear relationship. The blue line represents the initial linear function generated by the application of the minimum and maximum $\Delta_{\beta}$ values. The orange area shows the range for the bisection search. The black dot is the default rate point. The final rate point is the purple dot. It is within the maximum rate difference threshold and is closest to the default rate point.}
    \label{fig:LinearInter}
\end{figure}

As illustrated in Fig. \ref{fig:LinearInter}, in the preliminary phase of the search for $\Delta_{\beta}$, two constant values, $a$ and $b$, are calculated using the minimum and maximum values of $\Delta_{\beta}$ along with their corresponding rates. Then, we determine $\Delta_{\beta,1}$ by setting the linear function of ${\rm log}(R) = {\rm log}(R_t)$. We subsequently validate $\Delta_{\beta,1}$ in order to obtain the rate $R_1$. Thereafter, JPEG AI employs a bisection search for $\Delta_{\beta}$ in the range ${[\Delta_{\beta,1} - 100, \Delta_{\beta,1} + 100]}$. This search finds the final beta, which can generate a rate that is lower than the maximum rate difference threshold and is also closest to the current model's default rate. In this way, JPEG AI can match the rate and preserve the variable rate performance.

\begin{figure}
    \centering
    \includegraphics[width=0.48\textwidth]{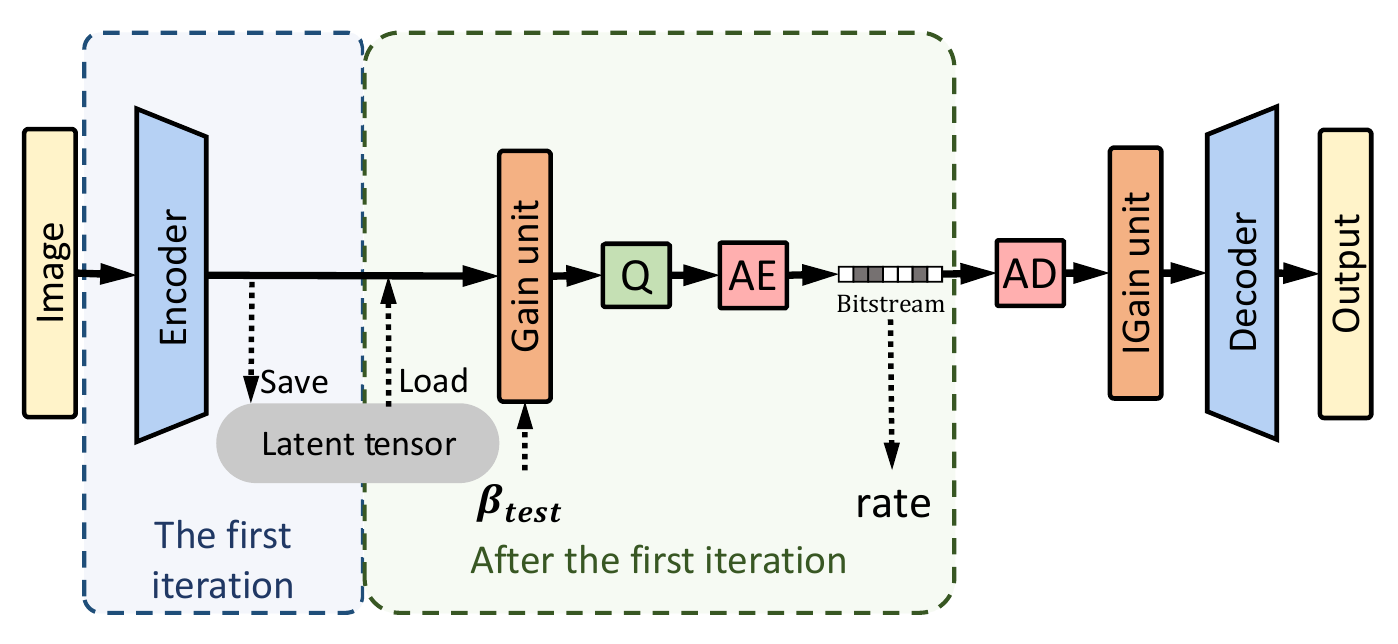}
    \caption{The illustration of $\Delta_{\beta}$ validation during the $\Delta_{\beta}$ search is presented here. The tensor before the gain unit is consistently identical in each $\Delta_{\beta}$ validation iteration, thus eliminating the necessity for repeated encoding.}
    \label{fig:validation}
\end{figure}
During the $\Delta_{\beta}$ search, each $\Delta_{\beta}$ need to be validated to get the real rate. Noting that $\Delta_{\beta}$ only affects the gain vector, which represents the channel-wise quantization map of the latent tensor, it follows that the latent tensor is unchanged before the gain vector is applied in each iteration. To prevent the repeated generation of the unmodified latent tensor, we suggest storing the unmodified latent tensor during the first iteration and retrieving it during following iterations. This approach saves computational time for the encoder. The $\Delta_{\beta}$ validation process is shown in Fig. \ref{fig:validation}.

In conclusion, the BRM algorithm in JPEG AI provides a solution to match the target rate with consideration of the variable rate feature in JPEG AI, thus avoiding redundant processes and ensuring an efficient algorithm capable of reaching the target rate.

\subsection{Variable Rate Training in JPEG AI VM}
In JPEG AI VM, the entire codec is trained in an end-to-end manner with a loss function that is identical to that described in \ref{eq_2}.  In order to encompass a more expansive range of rates, four models are trained using four distinct sets of $\beta_\mathrm{train}$ values. A higher value of $\beta_\mathrm{train}$ yields a more refined reconstructed image, which necessitates a greater number of bits for coding.
It should be noted that the CCS framework employs a separate processing approach for each color component. However, the different color component codecs are jointly trained with the same $\beta_\mathrm{train}$.

\begin{table*}
\centering
\caption{Variable Rate Training Strategy}
\label{tab:training}
\begin{tblr}{
  cell{1}{1} = {r=5}{},
  cell{2}{3} = {r=4}{},
  cell{3}{4} = {r=3}{},
  cell{7}{1} = {r=5}{},
  cell{8}{3} = {r=2}{},
  cell{9}{4} = {r=3}{},
  cell{10}{3} = {r=2}{},
  cell{13}{1} = {r=5}{},
  cell{14}{3} = {r=2}{},
  cell{15}{4} = {r=3}{},
  cell{16}{3} = {r=2}{},
  cell{19}{1} = {r=5}{},
  cell{20}{3} = {r=2}{},
  cell{21}{4} = {r=3}{},
  cell{22}{3} = {r=2}{},
  vline{2-6} = {1-2,7-8,13-14,19-20}{},
  vline{2-6,4-6} = {3,9,15,21}{},
  vline{2-6,6} = {4-5,11,17,23}{},
  vline{2-6,6} = {10,16,22}{},
  hline{1-2,6-8,12-14,18-20,24} = {-}{},
  hline{3,9,15,21} = {2,4-6}{},
  hline{4-5,11,17,23} = {2,5-6}{},
  hline{10,16,22} = {2-3,5-6}{},
  hline{11} = {2,6}{},
}
\textbf{Model 0} & \textbf{Training Stage} & $\bm{\beta_\mathrm{train}}$ & \textbf{Loss Type} & \textbf{Epochs} & \textbf{Trained Codec Parts}                                         \\
        & \RomanNumeralCaps{1}             & 0.002          & MSE       & 64     & encoder, decoder, entropy network \\
        & \RomanNumeralCaps{2}             &                & Mix       & 32     & encoder, decoder, entropy network \\
        & \RomanNumeralCaps{3}            &                &           & 20     & decoder, entropy network                   \\
        & \RomanNumeralCaps{4}             &                &           & 12     & gain unit                                            \\
        &                &                &           &        &                                                      \\
\textbf{Model 1} & \textbf{Training Stage} & $\bm{\beta_\mathrm{train}}$ & \textbf{Loss Type} & \textbf{Epochs} & \textbf{Trained Codec Parts}                                        \\
        & \RomanNumeralCaps{1}            & 0.007          & MSE       & 64     & encoder, decoder, entropy network \\
        & \RomanNumeralCaps{2}             &                & Mix      & 32     & encoder, decoder, entropy network \\
        & \RomanNumeralCaps{3}           & 0.03           &           & 20     & decoder, entropy network, gain unit    \\
        & \RomanNumeralCaps{4}             &                &           & 12     & decoder, gain unit                                \\
        &                &                &           &        &                                                      \\
\textbf{Model 2} & \textbf{Training Stage} & $\bm{\beta_\mathrm{train}}$ & \textbf{Loss Type} & \textbf{Epochs} & \textbf{Trained Codec Parts}                                         \\
        & \RomanNumeralCaps{1}              & 0.075          & MSE       & 64     & encoder, decoder, entropy network \\
        & \RomanNumeralCaps{2}             &                & Mix       & 32     & encoder, decoder, entropy network \\
        & \RomanNumeralCaps{3}            & 0.2~           &           & 20     & decoder, entropy network, gain unit    \\
        & \RomanNumeralCaps{4}             &                &           & 12     & decoder, gain unit                                \\
        &                &                &           &        &                                                      \\
\textbf{Model 3} & \textbf{Training Stage} & $\bm{\beta_\mathrm{train}}$ & \textbf{Loss Type} & \textbf{Epochs} & \textbf{Trained Codec Parts}                                         \\
        & \RomanNumeralCaps{1}              & 0.5            & MSE       & 64     & encoder, decoder, entropy network \\
        & \RomanNumeralCaps{2}             &                & Mix       & 32     & encoder, decoder, entropy network \\
        & \RomanNumeralCaps{3}            & 1              &           & 20     & decoder, entropy network, gain unit    \\
        & \RomanNumeralCaps{4}             &                &           & 12     & decoder, gain unit                                
\end{tblr}
\end{table*}

Table \ref{tab:training} illustrates the training strategy for four models. Each model will undergo training in four distinct stages, with the training parameters varying between stages. 

In general, at the training stage \RomanNumeralCaps{1}, the entire network will be trained using the MSE in the loss function, and training stage \RomanNumeralCaps{1} has the longest training time. In the training stage \RomanNumeralCaps{2}, the entire network will be trained using a mixed loss function. In this approach, the MSE and MS-SSIM losses are combined with different weights. In training stages \RomanNumeralCaps{3} and \RomanNumeralCaps{4}, a portion of the network is trained with the objective of fine-tuning the network and enabling variable rate adaptation.

It should be noted that the training strategy differs slightly between models. For model 0, which is designed to cover the low bit rate range, the $\beta_\mathrm{train}$ is identical across four stages. Additionally, fewer parts of model 0 are trained in stages \RomanNumeralCaps{3} and \RomanNumeralCaps{4} compared to other models. Furthermore, models 1, 2 and 3 utilise a larger $\beta_\mathrm{train}$ in training stages \RomanNumeralCaps{3} and \RomanNumeralCaps{4} than in training stages \RomanNumeralCaps{1} and \RomanNumeralCaps{2}. This is done for the purpose of enabling the models to adapt to a higher bit rate. However, in the evaluation, the $\beta_\mathrm{train}$ used to create $\Delta_{\beta}$ is consistent with the $\beta_\mathrm{train}$ used in training stage \RomanNumeralCaps{1}, as this stage has the greatest number of epochs and thus the model is most adapted to the $\beta_\mathrm{train}$ in this stage.
The aforementioned training strategy enables the adaptation of models to different quality levels through training with distinct $\beta_\mathrm{train}$ values. This approach facilitates the attainment of optimal variable rate performance during evaluation.

\section{Experiments}
\subsection{JPEG AI Common and Training Test Conditions}
The entire test of models is evaluated based on the JPEG AI common training and test condition (CTTC)~\cite{wg1n100106}. The training, validation, and the test dataset of CTTC are all in the PNG format with sRGB color space. The resolution of the JPEG AI dataset is from 256 $\times$ 256 to 8K pixels with 8 bits depth. And the JPEG AI training dataset has 5264 sequences, the validation dataset has 350 sequences, and the test dataset has 50 images, and the content of the JPEG AI dataset is natural and synthetic. When processing the PNG data, the input image will be transformed in YUV format according to BT.709 primaries. The internal process format in our experiment is set to the YUV444 format. 

To evaluate the performance, JPEG AI computed the average Bjøntegaard Delta rate (BD-Rate)~\cite{bjontegaard2001calculation} performance across 7 metrics, including MS-SSIM~\cite{1292216}, Visual Information Fidelity (VIF)~\cite{citVif}, feature similarity (FSIM)~\cite{5705575}, Normalized Laplacian Pyramid (NLPD)~\cite{nlpdcite}, Information Content Weighted Structural Similarity Measure (IW-SSIM)~\cite{5635337}, Video Multimethod Assessment Fusion (VMAF)~\cite{VMAFcite}, and PSNR-HVS-M~\cite{psnrHVS}.

When doing the BRM, JPEG AI CTTC set five target rate points to check the performance in different bit ranges, the target rates are 0.12, 0.25, 0.5, 0.75, and 1.0 bits per pixel (bpp).

\subsection{Channel-Wise Quality Map Performance}

This subsection presents the performance of using the channel-wise quality map to achieve overall variable rate. Initially, the variable rate performance of each model is demonstrated. Subsequently, the variable rate performance using BRM is illustrated, whereby BRM selects the appropriate model for the target rate. Finally, based on the CCS framework, the visual quality performance is exhibited using different channel-wise quality maps for luminance and chrominance color components.

\subsubsection{Variable Rate Curve of Each Model}

\begin{figure*}
    \centering
    \setkeys{Gin}{width=0.5\linewidth} 

\subfloat[\label{bop-msssim}]{\includegraphics[width=0.492\linewidth]{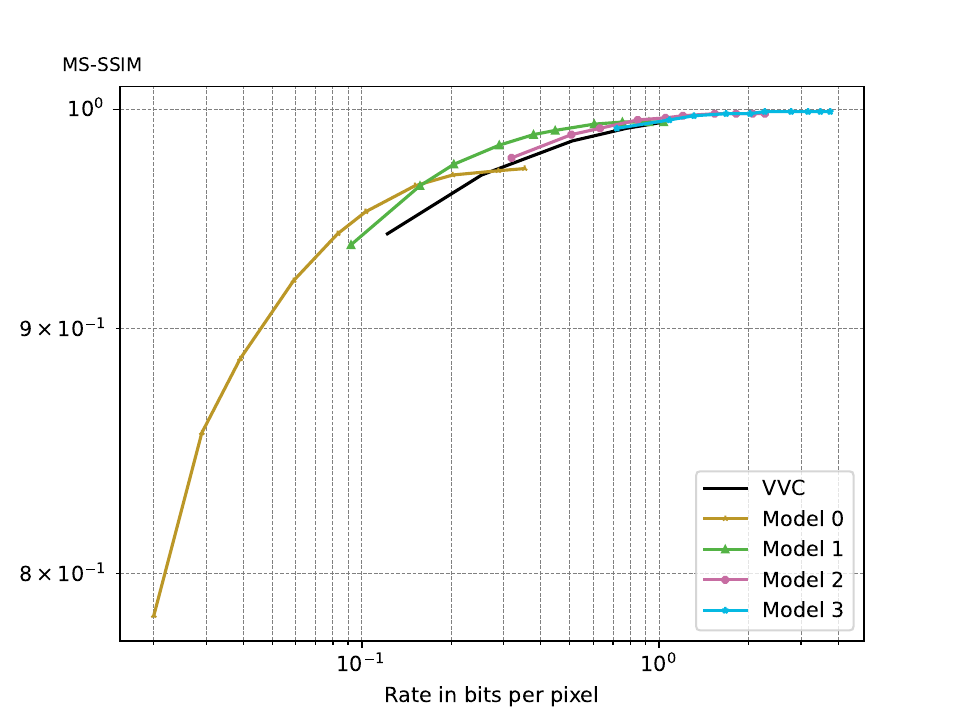} }\hfil
\subfloat[\label{bop-vamf}]{\includegraphics[width=0.492\linewidth]{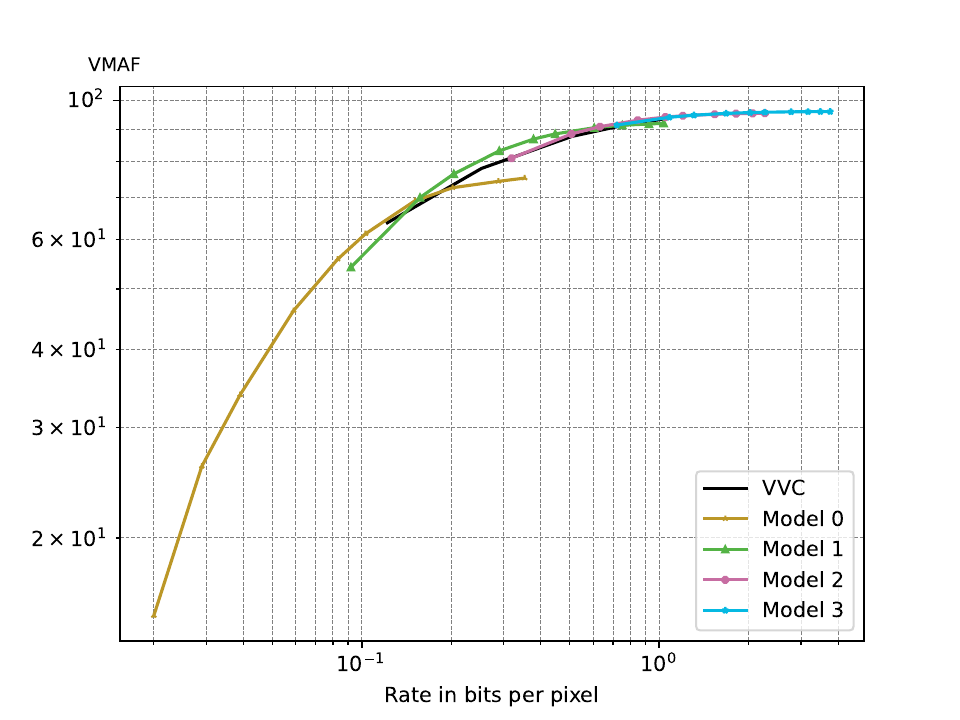} }\hfil

\caption{Variable rate curve of the main profile, the shown metrics are MS-SSIM and VMAF. The remaining 5 metrics are shown in the Appendix.}
\label{bop-vr}
\end{figure*}

For showing the variable rate performance of the each model, we use multiple $\Delta_{\beta}$ $ \{-1069, -860, -660, -460, -260, 0, 200, 400, 600, 702\}$ for each model. We used the same list of $\Delta_{\beta}$ values for each model to get the same scale of each $\beta_\mathrm{train}$. 


Fig. \ref{bop-vr} depicts the BD-rate curve of the main profile (MP), showcasing the performance of two metrics. Moreover, the VVC was situated at rate points of 0.12, 0.25, 0.5, 0.75, and 1.0 bpp as an anchor. The version of VVC that was utilized as the anchor is VTM 11.1. In order to facilitate clarity, the metrics value and rate axes are presented in logarithmic form. 

From Fig. \ref{bop-vr}, it can be observed that each model is capable of providing a limited bit rate range. Furthermore, within each provided bit rate range, a portion of the BD-rate curve of JPEG AI demonstrates superior performance in comparison to VVC. In certain instances, there are multiple models that can achieve a given rate value. In such cases, it is crucial to select an appropriate model. This is introduced in the section \ref{Sec_BRM}. As illustrated in Fig. \ref{bop-vr}, it can be observed that for models 2 and 3 at high rates,  an increase in the parameter $\Delta_{\beta}$ results in the metric reaching its saturation point, indicating that further gain is not possible.

\subsubsection{Bit Rate Matching Performance}
\begin{figure*}
    \centering
    \setkeys{Gin}{width=0.5\linewidth} 

\subfloat[\label{BRM-msssim}]{\includegraphics[width=0.492\linewidth]{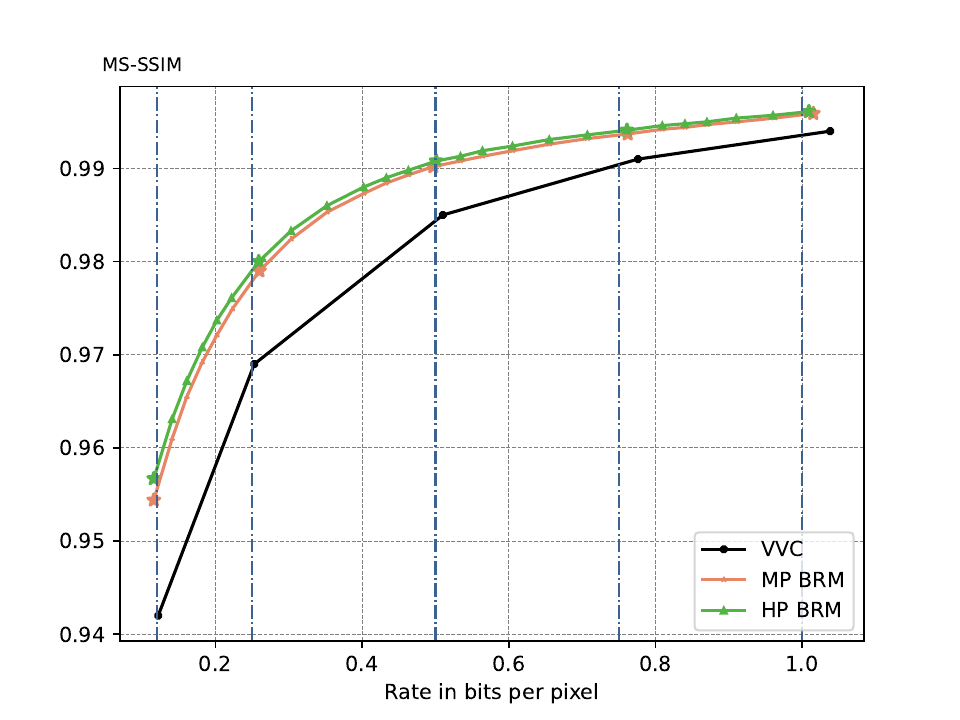} }\hfil
\subfloat[\label{BRM-vamf}]{\includegraphics[width=0.492\linewidth]{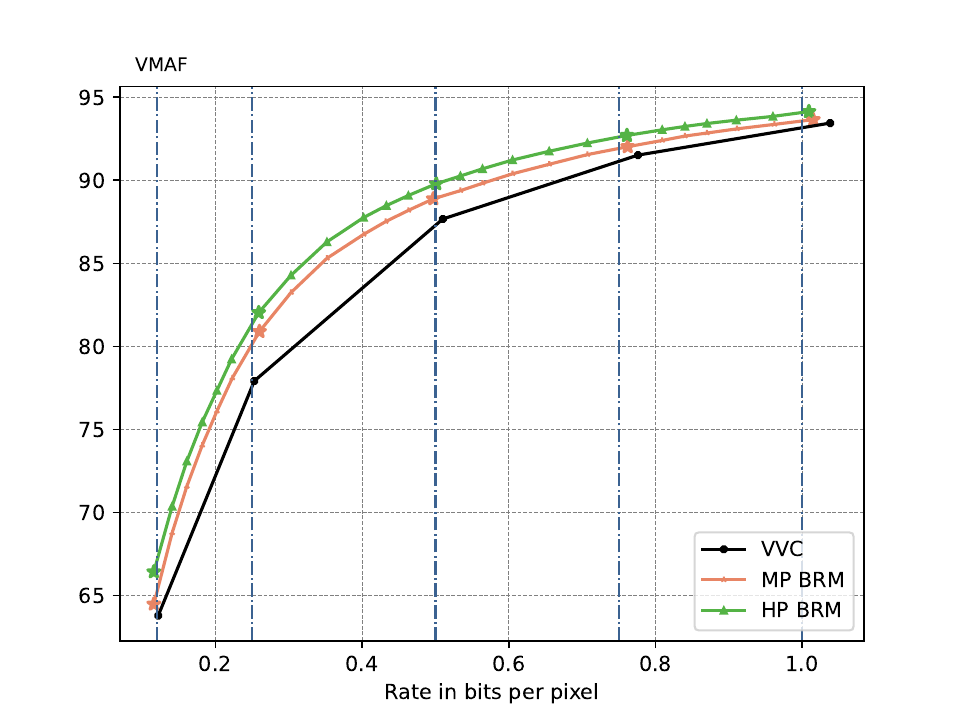} }\hfil

\caption{The figure illustrates the variable rate curve utilized for BRM. In addition to the target rate points, a series of rate points between two target rate points were plotted. The blue dashed lines indicate the target rate.the shown metrics are MS-SSIM and VMAF. }
\label{bop-brm}
\end{figure*}

In order to obtain the BRM result, two BD-rate curves are plotted, one for the main profile (MP) and one for the high profile (HP). The HP exhibits a higher level of complexity than the MP and is capable of performing better than the MP. The MS-SSIM and VMAF are the metrics used to evaluate the performance. The results are presented in Fig. \ref{BRM-msssim} and Fig. \ref{BRM-vamf}.

\begin{table*}[t]
        \newcommand\ph{\phantom{0}}
        \newcommand\phm{\phantom{-0}}

	\centering
	\caption{Average BD rate of BRM methods applied on MP and HP models of JPEG AI.}
	\label{resultTab}
	\footnotesize
	\begin{tabular}{l|c|c|c|c|c|c|c|c|c|c}
		\toprule
		\textbf{Model}&\textbf{AVG}&\textbf{ Max. Rate Diff.}&\textbf{Total Enc. Time}&\textbf{MSSSIM}&\textbf{VIF}&\textbf{FSIM}&\textbf{NLPD}&\textbf{IWSSIM}&\textbf{VMAF}&\textbf{PSNRHVS}\\
		\midrule
		\makecell[tl]{ VTM 11.1\\ VTM 23.4 \\ MP fixed rate \\MP  BRM \\ MP  BRM V2 \\ HP fixed rate \\ HP  BRM \\ HP BRM V2 }
		
		&\makecell[t]{\phm0.0\% \\ \ph-0.5\% \\ -16.7\%  \\  -13.1\%   \\ -12.6\%  \\ -24.0\%  \\ -19.2\%    \\ -18.7\%}
		&\makecell[t]{\ph10\%   \\ \ph10\%   \\  327\%   \\  \phm6\%   \\ \phm1\%  \\ 324\%    \\ \phm6\%    \\ \phm1\%}
        &\makecell[t]{  --      \\   --      \\ \ph5 min\\  15 min \\15 min   \\ 13 min\\ 67 min  \\ 66 min}
		&\makecell[t]{\phm0.0\% \\ \ph-0.9\% \\  -33.5\% \\ -29.3\%    \\ -28.8\%  \\ -38.2\%  \\ -33.4\%    \\ -32.9\%}
		&\makecell[t]{\phm0.0\% \\ \ph-0.5\% \\\phm0.8\% \\  \phm1.4\% \\ \phm1.7\%\\ \ph-9.1\%\\ \ph-6.8\%  \\ \ph-6.3\%}
		&\makecell[t]{\phm0.0\% \\ \ph-0.4\% \\  -20.2\% \\  -16.2\%   \\ -15.8\%  \\ -29.2\%  \\ -23.8\%    \\ -23.2\%}
		&\makecell[t]{\phm0.0\% \\ \ph-0.5\% \\  -16.2\% \\  -11.9\%   \\ -11.4\%  \\ -22.7\%  \\ -17.2\%    \\ -16.6\%}
		&\makecell[t]{\phm0.0\% \\ \ph-0.5\% \\  -29.2\% \\  -25.4\%   \\ -25.0\%  \\ -34.1\%  \\ -29.6\%    \\ -29.0\%}
		&\makecell[t]{\phm0.0\% \\ \ph-0.1\% \\  -15.1\% \\  -10.6\%   \\ -10.2\%\\ -24.0\%  \\ -18.1\%    \\ -17.5\%}
		&\makecell[t]{\phm0.0\% \\ \ph-0.8\% \\\ph-3.5\% \\  \phm0.5\% \\ \phm1.0\%\\ -10.7\%  \\ \ph-5.7\%  \\ \ph-5.1\%}
		
		\\
		\bottomrule
	\end{tabular}
\end{table*}

In the BRM test, the target rate was matched to 0.12, 0.25, 0.5, 0.75, and 1.0 bpp, with no more than a 10\% difference. As illustrated in Fig. \ref{BRM-msssim} and \ref{BRM-vamf}, MP and HP have demonstrated superior performance to that of VVC intra at equivalent rate points. Additionally, HP exhibits enhanced BD-rate performance compared to MP. For all quality metrics in JPEG AI CTTC, the BRM's performance is presented in Table \ref{resultTab}. All tests are run on a single Nvidia Titan RTX GPU. The maximum rate difference is the highest value from all the tests. It does not reflect the average value. The total time spent encoding for 250 sub-tests is shown by the total encoding time. This is because there are 50 images in the test data, and each image has five rate points. In a comparison with the fixed rate test of JPEG AI, BRM can help JPEG AI to match the rate within 10\% of the target rate. However, there is a drop in performance due to the variable rate coding, and the encoding time gets longer. Note that Table III shows two versions of the BRM test for JPEG AI. The BRM test V2 was set to match the target rate more closely. As previously stated, the BRM algorithm is not considered normative in the standardization. Implementers are encouraged to propose more efficient BRM solutions.  

\subsubsection{Different Quality Parameters to the Chrominance and Luminance Components}
JPEG AI has a dedicated framework for each colour component, resulting in the generation of two bitstreams. By assigning a distinct channel-wise quality map to each colour component, the JPEG AI model is capable of modulating the performance between luminance and chrominance. However, since all colour components are trained currently with the same quality parameters during the training phase, it is not straightforward to achieve superior average performance across all CTTC test sets by assigning different quality parameters to the individual colour components. This functionality is employed to assist with certain exceptional cases where users wish to enhance either luminance or chrominance performance by allocating additional bits.  

\begin{figure*}
    \centering
    \includegraphics[width=0.6\textwidth]{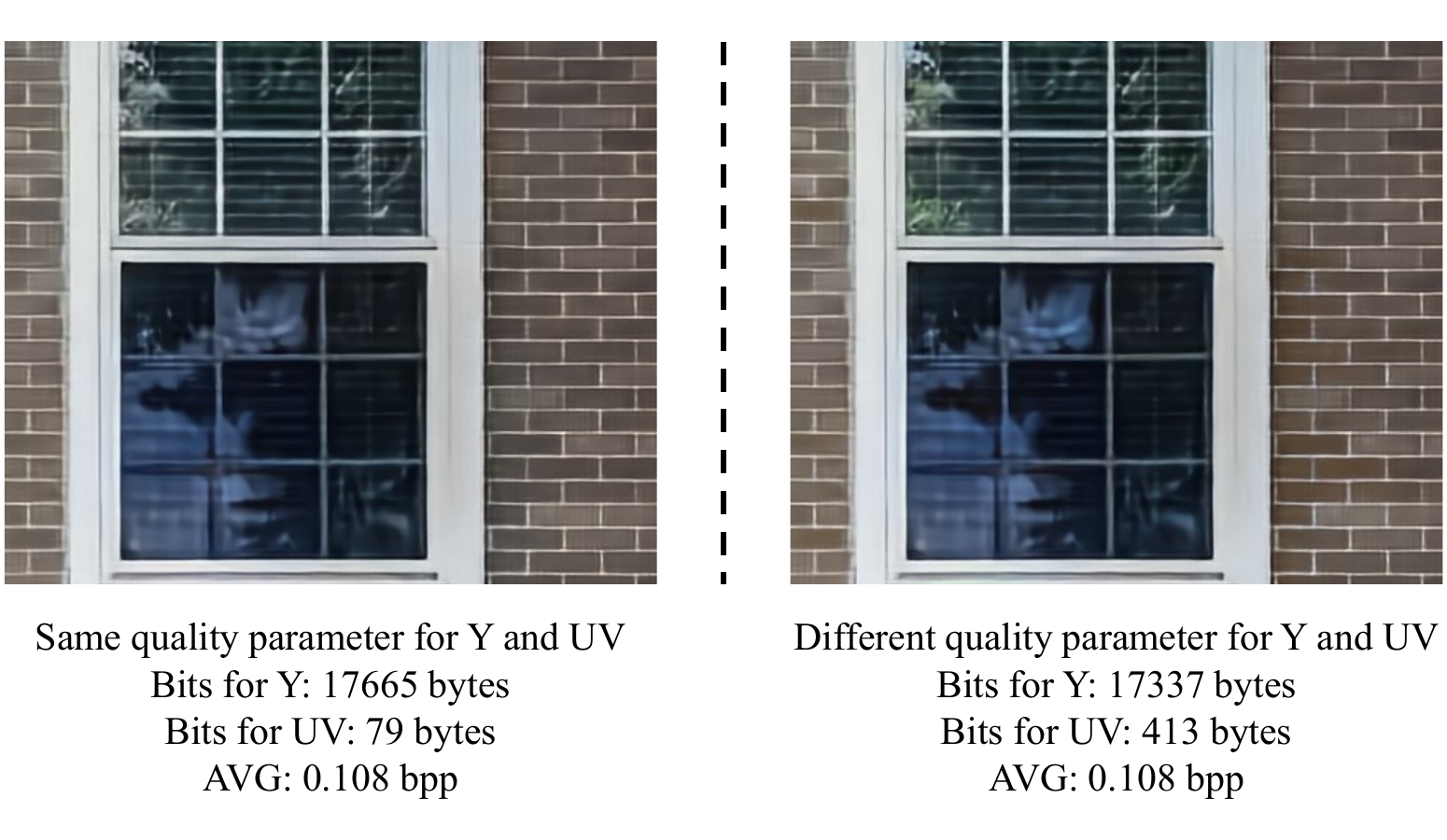}
    \caption{The figure illustrates the utilization of identical or disparate quality parameters for different color components. Both sides exhibit the same average rates; however, on the right side, a greater number of bits are allocated to chrominance components than on the left side. (The printed version may be of inferior visual quality. To obtain a clear visual representation of the result, please refer to the online version.)}
    \label{fig:BetaUV}
\end{figure*}

\begin{figure*}
    \centering
    \includegraphics[width=0.6\textwidth]{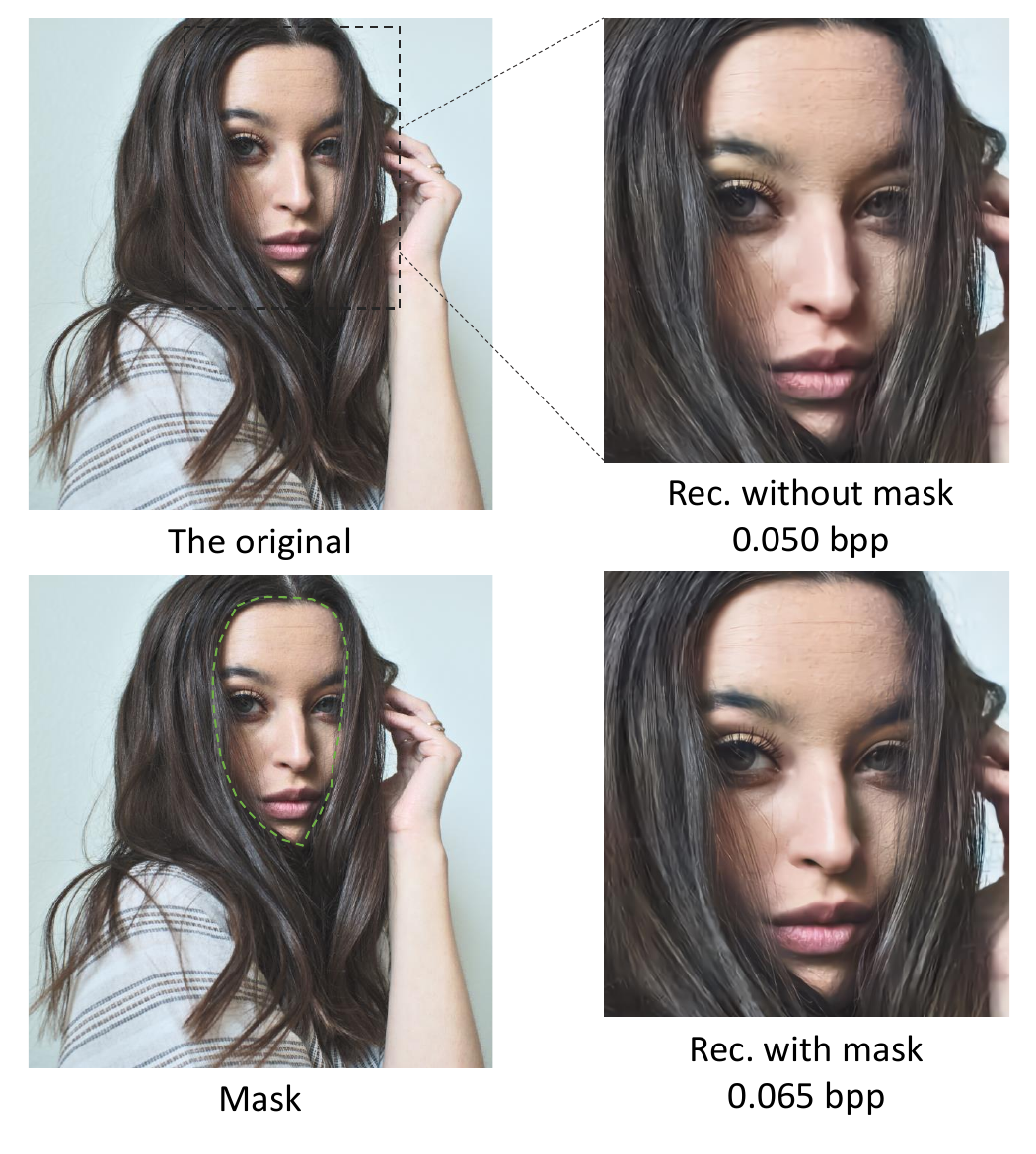}
    \caption{The illustration applying the spatial quality map on one woman's face, the mask is shown within the green dash line in the image. Masked area will be allocated more bits, in contrast, the rest area will be allocated less bits. In this example, more clear details can be found around the woman's face, and hair is more blurry. (The printed version may be of inferior visual quality. To obtain a clear visual representation of the result, please refer to the online version.)}
\label{fig:QP_map}
\end{figure*}

\begin{figure*}
    \centering
    \includegraphics[width=0.6\textwidth]{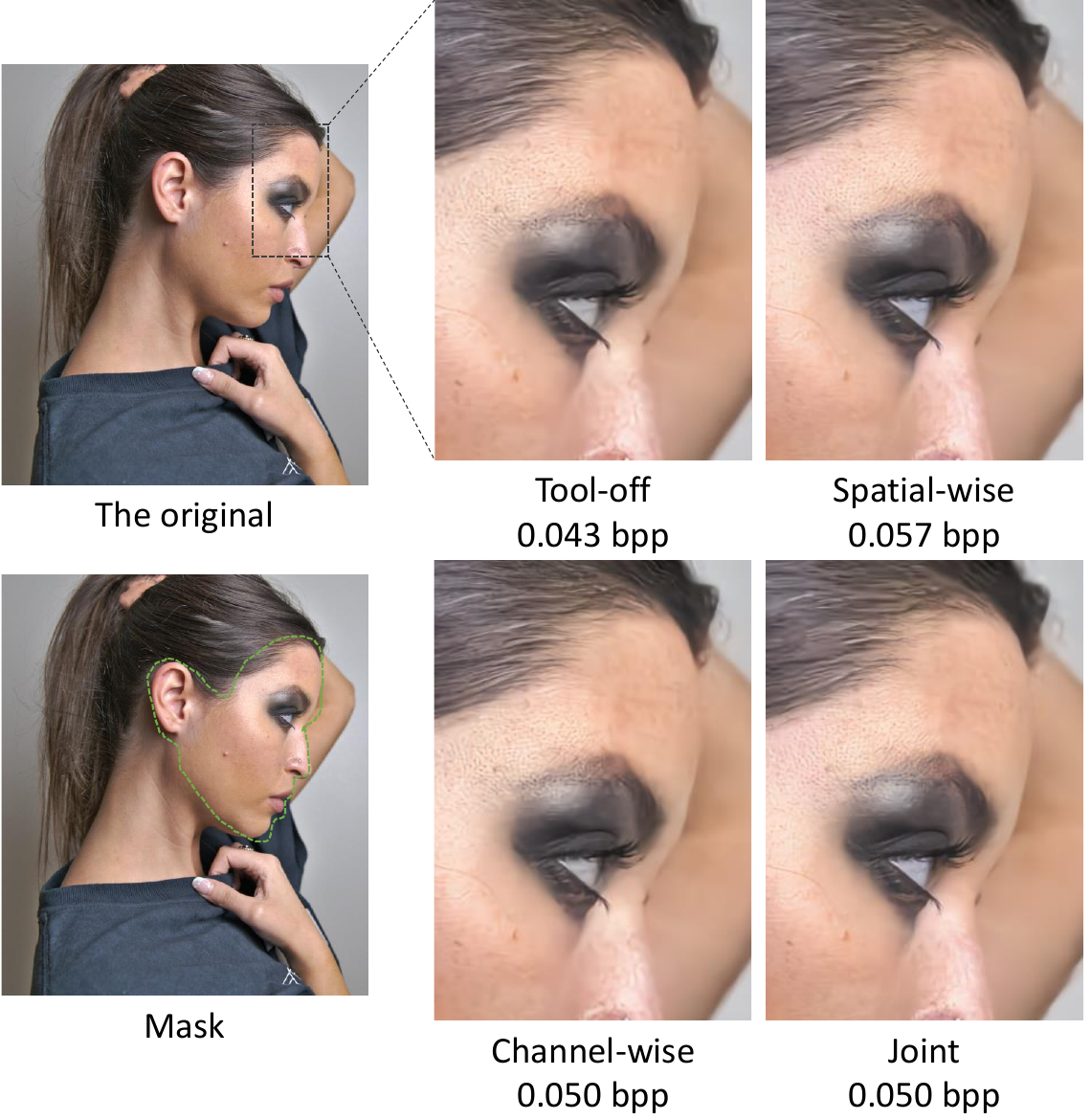}
    \caption{This illustration demonstrates the use of the spatial and channel-wise quality map, both separately and jointly. When utilising the channel-wise quality map, the rate was matched to 0.05 bpp through the application of BRM. Besides, the spatial quality map employs a ROI mask, which is represented within the green dash line in the image. (The printed version may be of inferior visual quality. To obtain a clear visual representation of the result, please refer to the online version.)}
    \label{fig:VisualJQ}
\end{figure*}
Fig. \ref{fig:BetaUV} illustrates the same portion of the image, employing identical and disparate quality parameters for the Y and UV color components. The left side of the figure utilizes the identical quality parameters, whereas the right side depicts the disparate quality parameters. Two distinct quality parameter assignments yielded an identical rate, however, on the right side of the figure, a slight reduction in bits for Y and an increase in bits for UV was implemented. The visual quality analysis reveals that the right side of the image is more colorful, as evidenced by the increased green in the trees and the more intense blue in the sky.

\subsection{Spatial Quality Map performance}
The spatial quality map provides the function for flexible allocating bits on different area, in the practical situation, this functionality is useful for the ROI coding. Thus, in this section, we shown the visual quality improvement of the ROI by applying the spatial quality map.

As shown in the Fig. \ref{fig:QP_map}, we applied the spatial quality map on the woman's face area. The test model is model 0 from the MP, we didn't apply the BRM on this image and just use $\delta_{\beta} = 1$ and a rate of 0.57 bpp. The spatial quality map is marked as the area enclosed by the green dotted line in the figure. To increase the bits that the codec spends on the ROI, we set the quality index 3 in the ROI. However, increasing the bits in the ROI can significantly increase the bits of the whole image. Thus, to avoid spending too many bits in the single image, we use quality index -3 for the non ROI. The result is shown in Fig. \ref{fig:QP_map}, where the women's hair is blurry and has less texture because it is located outside of the ROI. The original image has $1752\times1856$ pixels with size 5194KB. The test without applying the spatial quality map uses 0.50 bpp, and the test applying the quality map takes 0.65 bpp.

It can be concluded that the use of a spatial quality map allows for an improvement in ROI quality through the assignment of additional bits in the ROI. However, the application of a spatial quality map may result in a slight increase in the overall bit rate. To achieve enhanced ROI quality with a constant bit rate, the joint use of channel-wise and spatial quality maps can be employed. This experimental approach will be detailed in the subsequent subsection.

\subsection{Jointly using channel-wise and spatial quality map}
As illustrated in Fig. \ref{fig:JointQ}, the generation of a 3D quality map is achieved through the joint extension of a channel-wise and spatial quality map. This results in a ROI with a precisely defined bit rate.

Fig. \ref{fig:VisualJQ} illustrates the four reconstructed images. The model, trained with a $\beta_\mathrm{train}$ of 0.002, was employed to generate all test results. In the tool-off and spatial configuration, the channel-wise quality map was not modified, resulting in a $\delta_{\beta}$ of 1. In contrast, the channel-wise and joint configuration employed BRM to match 0.05 bpp.  From the test results, it can be observed that when the channel-wise and spatial quality maps are used in conjunction, the effects are cumulative. When the reconstructed images are compared, it can be seen that there is a notable difference between those that were only created using the channel-wise quality map and those that were created using a joint channel-wise and spatial quality map. It is possible to allocate a greater number of bits to the ROI while assigning fewer bits to the background. This can be achieved by incorporating a spatial quality map. However, the average rate remains unaltered due to the BRM function by tuning of the channel-wise quality map. In terms of visual quality, the ROI displays enhanced clarity in terms of skin and eye texture, while hair appears more blurred.

\section{Summary}
This paper introduces an overview of variable rate adaptations in the JPEG AI verification model. These include the variable rate training strategy, the three-dimensional quality map and the BRM algorithm. 
The verification model is trained with a training strategy that includes variable rate adaptation, whereby the channel-wise quality map is updated according to the quality parameters.

In the evaluation phase, a three-dimensional quality map is employed to facilitate variable rate adaptation. The three-dimensional quality map can be generated by extending either the channel-wise quality map generated from the quality parameters or the spatial quality map generated from ROI mask, or a combination thereof.

As the objective result, when applying the BRM, the main profile with low complexity yielded a 13\% BD-rate gain over VVC intra, while the high profile with high complexity achieved a 19.2\% BD-rate gain over VVC intra. With respect to the subjective results, the example of improving the quality of the ROI is illustrated.

In conclusion, this paper presents an overview of variable rate coding in JPEG AI, which enables flexible variable rate coding in a software- and hardware-friendly manner. The solutions introduced in this paper can be further applied to other learned-based image codecs.

\section*{Acknowledgments}
We would like to thank all the contributors from the International Electrotechnical Commission (IEC), the International Organization for Standardization (ISO) and the International Telecommunication Union (ITU). With their great help, JPEG AI could be successfully standardized.\\
Declaration of AI-assisted technologies: The authors used DeepL for grammar check and semantic rephrasing to improve the language.

{\appendix[Variable rate curve of the base operation point]

\begin{figure}
    \centering
    \includegraphics[width=0.45\textwidth]{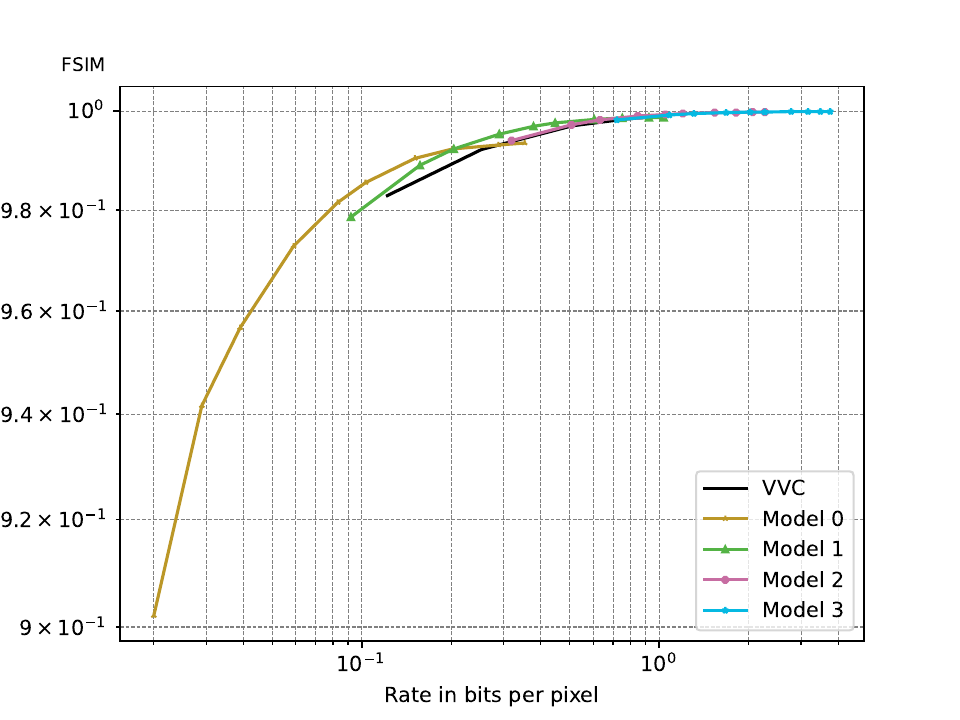}
    \caption{Variable rate curve of the main profile, the shown metric is FSIM.}
\end{figure}

\begin{figure}
    \centering
    \includegraphics[width=0.45\textwidth]{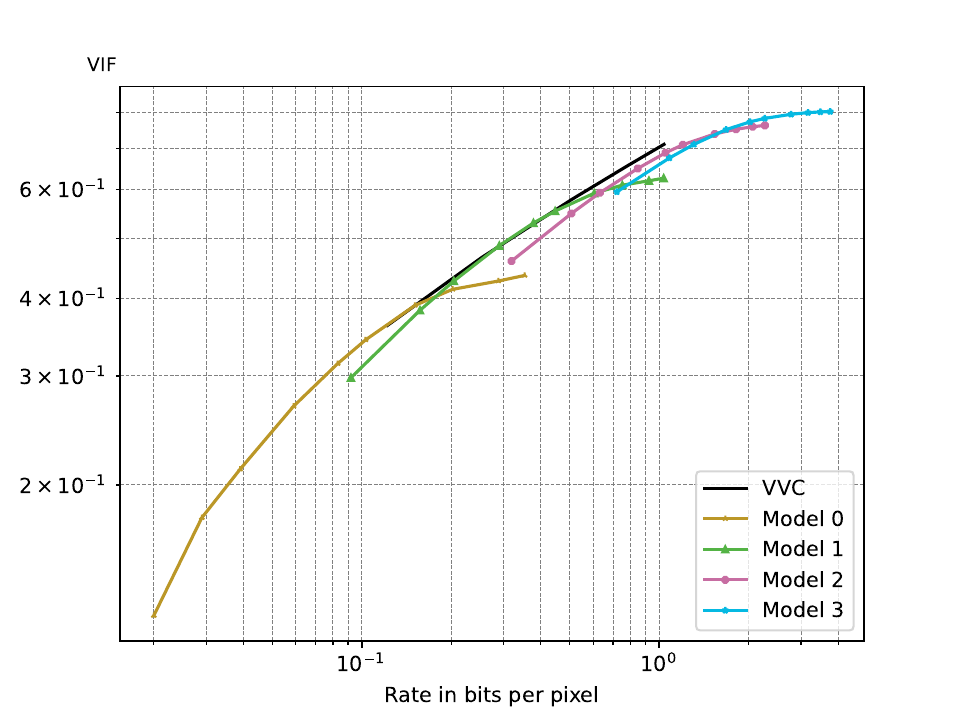}
    \caption{Variable rate curve of the main profile, the shown metric is VIF.}
\end{figure}

\begin{figure}
    \centering
    \includegraphics[width=0.45\textwidth]{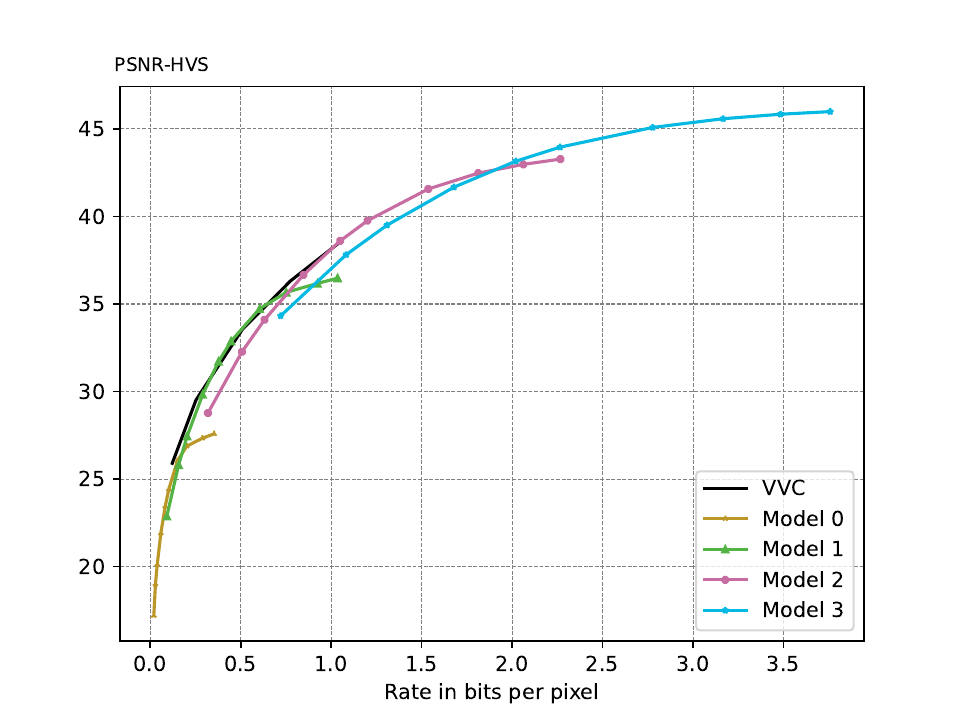}
    \caption{Variable rate curve of the main profile, the shown metric is PSNR-HVS.}
\end{figure}

\begin{figure}
    \centering
    \includegraphics[width=0.45\textwidth]{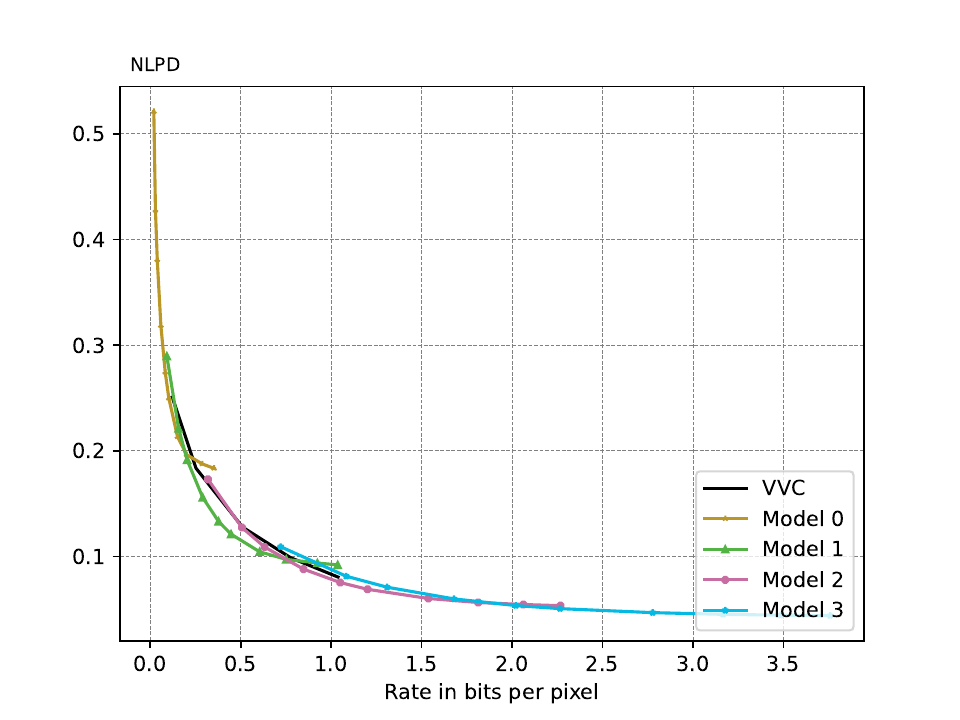}
    \caption{Variable rate curve of the main profile, the shown metric is NLPD.}
\end{figure}

\begin{figure}
    \centering
    \includegraphics[width=0.45\textwidth]{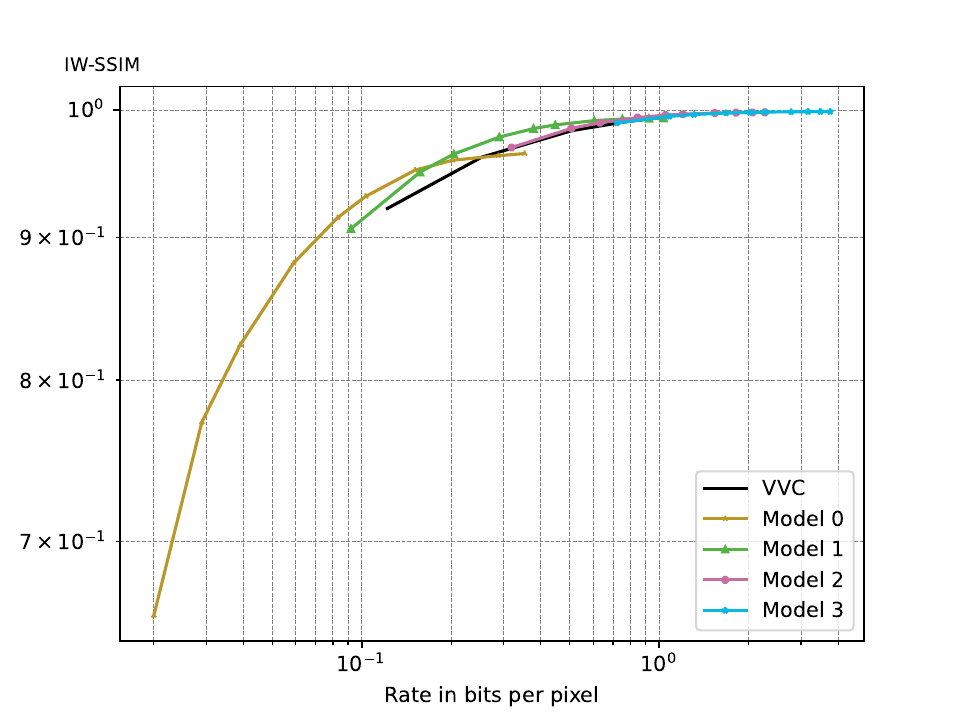}
    \caption{Variable rate curve of the main profile, the shown metric is IW-SSIM}
\end{figure}




\bibliographystyle{IEEEtran}
\bibliography{bib}

\begin{thebibliography}{10}
\providecommand{\url}[1]{#1}
\csname url@samestyle\endcsname
\providecommand{\newblock}{\relax}
\providecommand{\bibinfo}[2]{#2}
\providecommand{\BIBentrySTDinterwordspacing}{\spaceskip=0pt\relax}
\providecommand{\BIBentryALTinterwordstretchfactor}{4}
\providecommand{\BIBentryALTinterwordspacing}{\spaceskip=\fontdimen2\font plus
\BIBentryALTinterwordstretchfactor\fontdimen3\font minus \fontdimen4\font\relax}
\providecommand{\BIBforeignlanguage}[2]{{%
\expandafter\ifx\csname l@#1\endcsname\relax
\typeout{** WARNING: IEEEtran.bst: No hyphenation pattern has been}%
\typeout{** loaded for the language `#1'. Using the pattern for}%
\typeout{** the default language instead.}%
\else
\language=\csname l@#1\endcsname
\fi
#2}}
\providecommand{\BIBdecl}{\relax}
\BIBdecl

\bibitem{ma2019image}
S.~Ma, X.~Zhang, C.~Jia, Z.~Zhao, S.~Wang, and S.~Wang, ``Image and video compression with neural networks: A review,'' \emph{IEEE Transactions on Circuits and Systems for Video Technology}, vol.~30, no.~6, pp. 1683--1698, 2019.

\bibitem{birman2020overview}
R.~Birman, Y.~Segal, and O.~Hadar, ``Overview of research in the field of video compression using deep neural networks,'' \emph{Multimedia Tools and Applications}, pp. 1--24, 2020.

\bibitem{balle2017end}
J.~Ball{\'e}, V.~Laparra, and E.~P. Simoncelli, ``End-to-end optimized image compression,'' in \emph{5th International Conference on Learning Representations, ICLR 2017}, 2017.

\bibitem{balle2018variational}
J.~Ball{\'e}, D.~Minnen, S.~Singh, S.~J. Hwang, and N.~Johnston, ``Variational image compression with a scale hyperprior,'' in \emph{International Conference on Learning Representations}, 2018.

\bibitem{minnen2018joint}
D.~Minnen, J.~Ball{\'e}, and G.~Toderici, ``Joint autoregressive and hierarchical priors for learned image compression,'' in \emph{NeurIPS}, 2018.

\bibitem{liu2019non}
T.~Chen, H.~Liu, Z.~Ma, Q.~Shen, X.~Cao, and Y.~Wang, ``End-to-end learnt image compression via non-local attention optimization and improved context modeling,'' \emph{IEEE Transactions on Image Processing}, vol.~30, pp. 3179--3191, 2021.

\bibitem{zhou2019multi}
J.~Zhou, S.~Wen, A.~Nakagawa, K.~Kazui, and Z.~Tan, ``Multi-scale and context-adaptive entropy model for image compression,'' \emph{arXiv preprint arXiv:1910.07844}, 2019.

\bibitem{lee2018context}
J.~Lee, S.~Cho, and S.-K. Beack, ``Context-adaptive entropy model for end-to-end optimized image compression,'' in \emph{6th International Conference on Learning Representations, ICLR 2018}, 2018.

\bibitem{mentzer2018conditional}
F.~Mentzer, E.~Agustsson, M.~Tschannen, R.~Timofte, and L.~Van~Gool, ``Conditional probability models for deep image compression,'' in \emph{Proceedings of the IEEE Conference on Computer Vision and Pattern Recognition}, 2018, pp. 4394--4402.

\bibitem{cheng2020learned}
Z.~Cheng, H.~Sun, M.~Takeuchi, and J.~Katto, ``Learned image compression with discretized gaussian mixture likelihoods and attention modules,'' in \emph{Proceedings of the IEEE/CVF Conference on Computer Vision and Pattern Recognition}, 2020, pp. 7939--7948.

\bibitem{qian2020learning}
\BIBentryALTinterwordspacing
Y.~Qian, Z.~Tan, X.~Sun, M.~Lin, D.~Li, Z.~Sun, H.~Li, and R.~Jin, ``Learning accurate entropy model with global reference for image compression,'' 2022. [Online]. Available: \url{https://arxiv.org/abs/2010.08321}
\BIBentrySTDinterwordspacing

\bibitem{9067005}
M.~Li, K.~Ma, J.~You, D.~Zhang, and W.~Zuo, ``Efficient and effective context-based convolutional entropy modeling for image compression,'' \emph{IEEE Transactions on Image Processing}, vol.~29, pp. 5900--5911, 2020.

\bibitem{koyuncu2021parallel}
A.~B. Koyuncu, K.~Cui, A.~Boev, and E.~Steinbach, ``Parallelized context modeling for faster image coding,'' in \emph{2021 International Conference on Visual Communications and Image Processing (VCIP)}.\hskip 1em plus 0.5em minus 0.4em\relax IEEE, 2021, pp. 1--5.

\bibitem{he2021checkerboard}
D.~He, Y.~Zheng, B.~Sun, Y.~Wang, and H.~Qin, ``Checkerboard context model for efficient learned image compression,'' in \emph{Proceedings of the IEEE/CVF Conference on Computer Vision and Pattern Recognition}, 2021, pp. 14\,771--14\,780.

\bibitem{minnen2020channel}
D.~Minnen and S.~Singh, ``Channel-wise autoregressive entropy models for learned image compression,'' in \emph{2020 IEEE International Conference on Image Processing (ICIP)}.\hskip 1em plus 0.5em minus 0.4em\relax IEEE, 2020, pp. 3339--3343.

\bibitem{qian2021entroformer}
Y.~Qian, X.~Sun, M.~Lin, Z.~Tan, and R.~Jin, ``Entroformer: A transformer-based entropy model for learned image compression,'' in \emph{International Conference on Learning Representations}, 2021.

\bibitem{li2021deepcontextualvideocompression}
\BIBentryALTinterwordspacing
J.~Li, B.~Li, and Y.~Lu, ``Deep contextual video compression,'' 2021. [Online]. Available: \url{https://arxiv.org/abs/2109.15047}
\BIBentrySTDinterwordspacing

\bibitem{Ladune2020OpticalFA}
T.~Ladune, P.~Philippe, W.~Hamidouche, L.~Zhang, and O.~D{\'e}forges, ``{Optical Flow and Mode Selection for Learning-based Video Coding},'' \emph{Proc. IEEE 22nd International Workshop on Multimedia Signal Processing (MMSP)}, pp. 1--6, 2020.

\bibitem{jia2022learningbased}
P.~Jia, A.~B. Koyuncu, G.~Gaikov, A.~Karabutov, E.~Alshina, and A.~Kaup, ``Learning-based conditional image coder using color separation,'' pp. 49--53, 2022.

\bibitem{Choi2019VariableRD}
Y.~Choi, M.~El-Khamy, and J.~Lee, ``{Variable Rate Deep Image Compression With a Conditional Autoencoder},'' \emph{2019 IEEE/CVF International Conference on Computer Vision (ICCV)}, pp. 3146--3154, 2019.

\bibitem{song2021variable}
M.~Song, J.~Choi, and B.~Han, ``Variable-rate deep image compression through spatially-adaptive feature transform,'' in \emph{Proceedings of the IEEE/CVF International Conference on Computer Vision}, 2021, pp. 2380--2389.

\bibitem{9522770}
F.~Brand, K.~Fischer, and A.~Kaup, ``Rate-distortion optimized learning-based image compression using an adaptive hierachical autoencoder with conditional hyperprior,'' in \emph{2021 IEEE/CVF Conference on Computer Vision and Pattern Recognition Workshops (CVPRW)}, 2021, pp. 1885--1889.

\bibitem{Cui2020GVAEAC}
\BIBentryALTinterwordspacing
Z.~Cui, J.~Wang, B.~Bai, T.~Guo, and Y.~Feng, ``G-vae: A continuously variable rate deep image compression framework,'' \emph{ArXiv}, vol. abs/2003.02012, 2020. [Online]. Available: \url{https://api.semanticscholar.org/CorpusID:211988449}
\BIBentrySTDinterwordspacing

\bibitem{cui2021asymmetric}
Z.~Cui, J.~Wang, S.~Gao, T.~Guo, Y.~Feng, and B.~Bai, ``Asymmetric gained deep image compression with continuous rate adaptation,'' in \emph{Proceedings of the IEEE/CVF Conference on Computer Vision and Pattern Recognition}, 2021, pp. 10\,532--10\,541.

\bibitem{wang2023evcrealtimeneuralimage}
\BIBentryALTinterwordspacing
G.-H. Wang, J.~Li, B.~Li, and Y.~Lu, ``Evc: Towards real-time neural image compression with mask decay,'' 2023. [Online]. Available: \url{https://arxiv.org/abs/2302.05071}
\BIBentrySTDinterwordspacing

\bibitem{10274142}
Z.~Duan, M.~Lu, J.~Ma, Y.~Huang, Z.~Ma, and F.~Zhu, ``Qarv: Quantization-aware resnet vae for lossy image compression,'' \emph{IEEE Transactions on Pattern Analysis and Machine Intelligence}, vol.~46, no.~1, pp. 436--450, 2024.

\bibitem{VR_side}
Y.~Shi, K.~Zhang, J.~Wang, N.~Ling, and B.~Yin, ``Variable-rate image compression based on side information compensation and r $-\lambda$ model rate control,'' \emph{IEEE Transactions on Circuits and Systems for Video Technology}, vol.~PP, pp. 1--1, 01 2022.

\bibitem{wallace1992jpeg}
G.~K. Wallace, ``The jpeg still picture compression standard,'' \emph{IEEE Transactions on Consumer Electronics}, vol.~38, no.~1, pp. xviii--xxxiv, 1992.

\bibitem{skodras2001jpeg}
A.~Skodras, C.~Christopoulos, and T.~Ebrahimi, ``The jpeg 2000 still image compression standard,'' \emph{IEEE Signal Processing Magazine}, vol.~18, no.~5, pp. 36--58, 2001.

\bibitem{bellard2015bpg}
F.~Bellard, ``Bpg image format,'' 2015, \url{https://bellard.org/bpg} (accessed Dec. 01, 2022).

\bibitem{sze2014high}
V.~Sze, M.~Budagavi, and G.~J. Sullivan, ``High efficiency video coding (hevc),'' in \emph{Integrated Circuit and Systems, Algorithms and Architectures}.\hskip 1em plus 0.5em minus 0.4em\relax Springer, 2014, vol.~39, p.~40.

\bibitem{he2022elic}
D.~He, Z.~Yang, W.~Peng, R.~Ma, H.~Qin, and Y.~Wang, ``Elic: Efficient learned image compression with unevenly grouped space-channel contextual adaptive coding,'' in \emph{Proceedings of the IEEE/CVF Conference on Computer Vision and Pattern Recognition}, 2022, pp. 5718--5727.

\bibitem{guo2021causal}
G.~Zongyu, Z.~Zhang, R.~Feng, and Z.~Chen, ``Causal contextual prediction for learned image compression,'' \emph{IEEE Transactions on Circuits and Systems for Video Technology}, vol.~PP, pp. 1--1, 06 2021.

\bibitem{koyuncu2022contextformer}
A.~B. Koyuncu, H.~Gao, A.~Boev, G.~Gaikov, E.~Alshina, and E.~Steinbach, ``Contextformer: A transformer with spatio-channel attention for context modeling in learned image compression,'' in \emph{Computer Vision--ECCV 2022: 17th European Conference, Tel Aviv, Israel, October 23--27, 2022, Proceedings, Part XIX}.\hskip 1em plus 0.5em minus 0.4em\relax Springer, 2022, pp. 447--463.

\bibitem{effcient_context}
A.~B. Koyuncu, P.~Jia, A.~Boev, E.~Alshina, and E.~Steinbach, ``Efficient contextformer: Spatio-channel window attention for fast context modeling in learned image compression,'' \emph{IEEE Transactions on Circuits and Systems for Video Technology}, vol.~PP, pp. 1--1, 08 2024.

\bibitem{ohm2018versatile}
``{Versatile Video Coding},'' Rec. ITU-T H.266 and ISO/IEC 23090-3, Standard, Aug. 2020.

\bibitem{ascenso2021white}
J.~Ascenso and E.~Upenik, ``White paper on jpeg ai scope and framework v1. 0,'' \emph{ISO/IEC JTC 1/SC 29/WG1 N90049}, 2021.

\bibitem{wg1n83058}
\BIBentryALTinterwordspacing
J.~Ascenso and P.~Akayzi, ``{Report on the State-of-the-art of Learning based Image Coding },'' ISO/IEC JTC 1/SC 29/WG 1, Tech. Rep. wg1n83058, 2019. [Online]. Available: \url{https://jpeg.org/downloads/jpegai/wg1n83058-learning_based_image_coding_report.pdf}
\BIBentrySTDinterwordspacing

\bibitem{wg1n86018}
\BIBentryALTinterwordspacing
ISO/IEC and JTC1/SC29/WG1, ``{Call for Evidence on Learning-based Image Coding Technologies (JPEG AI) },'' ISO/IEC JTC 1/SC 29/WG 1, Tech. Rep. wg1n86018, 2020. [Online]. Available: \url{https://ds.jpeg.org/documents/jpegai/wg1n86018-REQ-CfE_JPEG_AI.pdf}
\BIBentrySTDinterwordspacing

\bibitem{wg1n89022}
\BIBentryALTinterwordspacing
J.~Ascenso, E.~Upenik, M.~Testolina, E.~Alshina, A.~Boev, and N.~Giuliani, ``{Report on the JPEG AI Call for Evidence Results },'' ISO/IEC JTC 1/SC 29/WG 1, Tech. Rep. wg1n89022, 2020. [Online]. Available: \url{https://ds.jpeg.org/documents/jpegai/wg1n89022-REQ-Report_on_the_JPEG_AI_CfE_Results.pdf}
\BIBentrySTDinterwordspacing

\bibitem{wg1n100095}
\BIBentryALTinterwordspacing
ISO/IEC and JTC1/SC29/WG1, ``{Final Call for Proposals for JPEG AI},'' ISO/IEC JTC 1/SC 29/WG 1, Tech. Rep. wg1n100095, 2022. [Online]. Available: \url{https://ds.jpeg.org/documents/jpegai/wg1n100095-094-REQ-Final_Call_for_Proposals_for_JPEG_AI.pdf}
\BIBentrySTDinterwordspacing

\bibitem{wg1n100250}
\BIBentryALTinterwordspacing
REQ, ``{Report on the JPEG AI Call for Proposals Results},'' ISO/IEC JTC 1/SC 29/WG 1, Tech. Rep. wg1n100250, 2022. [Online]. Available: \url{https://ds.jpeg.org/documents/jpegai/wg1n100250-096-REQ-Report_on_the_JPEG_AI_Call_for_Proposals_Results.pdf}
\BIBentrySTDinterwordspacing

\bibitem{wg1n100106}
\BIBentryALTinterwordspacing
ISO/IEC and JTC1/SC29/WG1, ``{JPEG AI Common Training and Test Conditions},'' ISO/IEC JTC 1/SC 29/WG 1, Tech. Rep. wg1n100106, 2022. [Online]. Available: \url{https://ds.jpeg.org/documents/jpegai/wg1n100106-094-ICQ-JPEG_AI_Common_Training_and_Test_Conditions.pdf}
\BIBentrySTDinterwordspacing

\bibitem{wg1n100279}
\BIBentryALTinterwordspacing
ICQ, ``{Description of the JPEG AI Verification Model under Consideration and associated software integration procedure },'' ISO/IEC JTC 1/SC 29/WG 1, Tech. Rep. wg1n100279, 2022. [Online]. Available: \url{https://ds.jpeg.org/documents/jpegai/wg1n100279-096-ICQ-Description_of_the_JPEG_AI_Verification_Model_under_Consideration_and_associated_software_integration_procedure.pdf}
\BIBentrySTDinterwordspacing

\bibitem{wg1n100658}
\BIBentryALTinterwordspacing
JTC1/SC29/WG1, ``{JPEG AI Overview Slides},'' ISO/IEC JTC 1/SC 29/WG 1, Tech. Rep. wg1n100658, 2023. [Online]. Available: \url{https://ds.jpeg.org/documents/jpegai/wg1n100658-101-COM-JPEG_AI_Overview_Slides.zip}
\BIBentrySTDinterwordspacing

\bibitem{jia2024bitdistributionstudyimplementation}
\BIBentryALTinterwordspacing
P.~Jia, J.~Mao, E.~Koyuncu, A.~B. Koyuncu, T.~Solovyev, A.~Karabutov, Y.~Zhao, E.~Alshina, and A.~Kaup, ``Bit distribution study and implementation of spatial quality map in the jpeg-ai standardization,'' 2024. [Online]. Available: \url{https://arxiv.org/abs/2402.17470}
\BIBentrySTDinterwordspacing

\bibitem{1292216}
Z.~Wang, E.~Simoncelli, and A.~Bovik, ``Multiscale structural similarity for image quality assessment,'' in \emph{The Thrity-Seventh Asilomar Conference on Signals, Systems \& Computers, 2003}, vol.~2, 2003, pp. 1398--1402 Vol.2.

\bibitem{citVif}
H.~Sheikh and A.~Bovik, ``Image information and visual quality,'' vol.~3, 01 2004, pp. iii--709.

\bibitem{5705575}
L.~Zhang, L.~Zhang, X.~Mou, and D.~Zhang, ``Fsim: A feature similarity index for image quality assessment,'' \emph{IEEE Transactions on Image Processing}, vol.~20, no.~8, pp. 2378--2386, 2011.

\bibitem{nlpdcite}
V.~Laparra, J.~Ballé, A.~Berardino, and E.~Simoncelli, ``Perceptual image quality assessment using a normalized laplacian pyramid,'' \emph{Electronic Imaging}, vol. 2016, pp. 1--6, 02 2016.

\bibitem{5635337}
Z.~Wang and Q.~Li, ``Information content weighting for perceptual image quality assessment,'' \emph{IEEE Transactions on Image Processing}, vol.~20, no.~5, pp. 1185--1198, 2011.

\bibitem{VMAFcite}
Z.~Li, A.~Aaronand, I.~Katsavounidis, A.~Moorthy, and M.~Manohara, ``{Toward A Practical Perceptual Video Quality Metric},'' 2016, \url{https://netflixtechblog.com/toward-a-practical-perceptual-video-quality-metric-653f208b9652}.

\bibitem{psnrHVS}
N.~Ponomarenko, F.~Silvestri, K.~Egiazarian, M.~Carli, J.~Astola, and V.~Lukin, ``On between-coefficient contrast masking of dct basis functions,'' \emph{Proc of the 3rd Int Workshop on Video Processing and Quality Metrics for Consumer Electronics}, 01 2007.

\bibitem{9244548}
F.~Brand, J.~Seiler, and A.~Kaup, ``{Intra-Frame Coding Using a Conditional Autoencoder},'' \emph{IEEE Journal of Selected Topics in Signal Processing}, vol.~15, no.~2, pp. 354--365, 2021.

\bibitem{10247017}
Z.~Zhang, S.~Esenlik, Y.~Wu, M.~Wang, K.~Zhang, and L.~Zhang, ``End-to-end learning-based image compression with a decoupled framework,'' \emph{IEEE Transactions on Circuits and Systems for Video Technology}, vol.~34, no.~5, pp. 3067--3081, 2024.

\bibitem{10448359}
E.~Koyuncu, T.~Solovyev, J.~Sauer, E.~Alshina, and A.~Kaup, ``Quantized decoder in learned image compression for deterministic reconstruction,'' in \emph{ICASSP 2024 - 2024 IEEE International Conference on Acoustics, Speech and Signal Processing (ICASSP)}, 2024, pp. 3985--3989.

\bibitem{10018040}
E.~Koyuncu, T.~Solovyev, E.~Alshina, and A.~Kaup, ``Device interoperability for learned image compression with weights and activations quantization,'' in \emph{2022 Picture Coding Symposium (PCS)}, 2022, pp. 151--155.

\bibitem{faucris.325748200}
P.~Jia, A.~B. Koyuncu, J.~Mao, Z.~Cui, Y.~Ma, T.~Guo, T.~Solovyev, A.~Karabutov, Y.~Zhao, J.~Wang, E.~Alshina, and A.~Kaup, ``{Bit} {Rate} {Matching} {Algorithm} {Optimization} in {JPEG}-{AI} {Verification} {Model},'' in \emph{2024 Picture Coding Symposium, PCS 2024 - Proceedings}.\hskip 1em plus 0.5em minus 0.4em\relax Institute of Electrical and Electronics Engineers Inc., 2024, cRIS-Team Scopus Importer:2024-07-19.

\bibitem{bjontegaard2001calculation}
\BIBentryALTinterwordspacing
G.~Bj{\o}ntegaard., ``Calculation of average psnr differences between rd-curves,'' \emph{ITU-T VCEG-M33}, 2001. [Online]. Available: \url{https://cir.nii.ac.jp/crid/1572543025125831168}
\BIBentrySTDinterwordspacing

\end{thebibliography}
\begin{IEEEbiography}[{\includegraphics[width=1in,height=1.25in,clip,keepaspectratio]{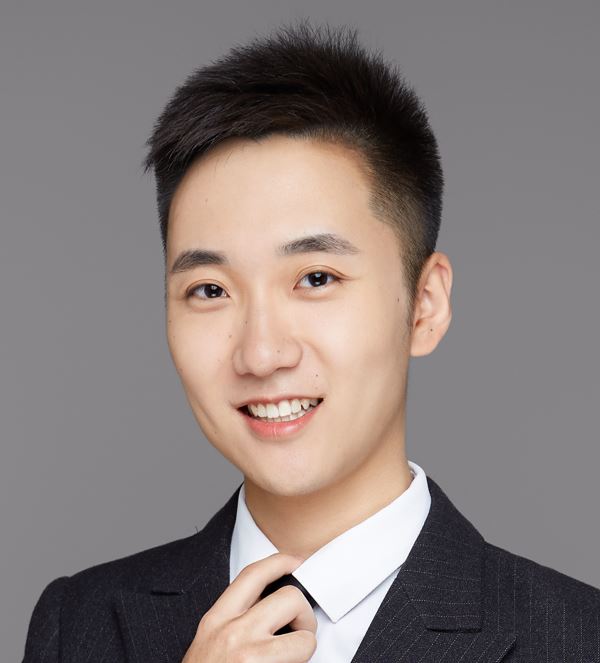}}]{Panqi Jia}
obtained his master’s degree in electrical engineering from Leibniz Universität Hannover (LUH) in Hannover, Germany in 2019. During his undergraduate studies, he conducted research on the application of ultra-wideband radar. During his master's degree, he conducted research into Car to X (C2X) communication systems and stereo matching algorithms within the field of vehicle communication. Since 2020, he has been a joint PhD student at Huawei Technologies, which has collaborated with the Chair of Multimedia Communications and Signal Processing at Friedrich Alexander University (FAU). He has been instrumental in the development of the JPEG AI standardization since 2021, acting as the core experiment coordinator in several rounds. His research activities include the investigation of methods for image and video compression and deep learning.
\end{IEEEbiography}
\begin{IEEEbiography}[{\includegraphics[width=1in,height=1.25in,clip,keepaspectratio]{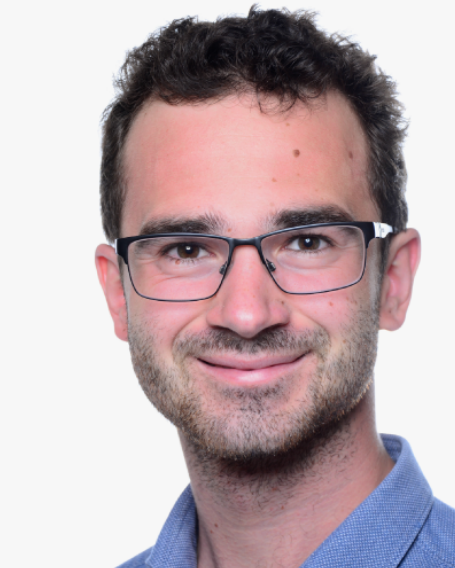}}]{Fabian Brand}
received his PhD in electrical engineering from Friedrich-Alexander Universität Erlangen-Nürnberg (FAU), Germany, in 2024. His main research interest is image and video compression using learning-based techniques. During his bachelor's, he worked on methods for frame-rate-conversion of video sequences, and during his master's, he researched automated harmonic analysis of classical music and style classification. From 2019 to 2023, he has been a Researcher with the Chair of Multimedia Communications and Signal Processing, FAU, where he conducts research on methods for video compression and deep learning, before joining the Huawei Munich Research Center to continue his work in compression. For his work, among others, he received the Best Paper Award of the Picture Coding Symposium (PCS) 2019.
\end{IEEEbiography}
\begin{IEEEbiography}[{\includegraphics[width=1in,height=1.25in,clip,keepaspectratio]{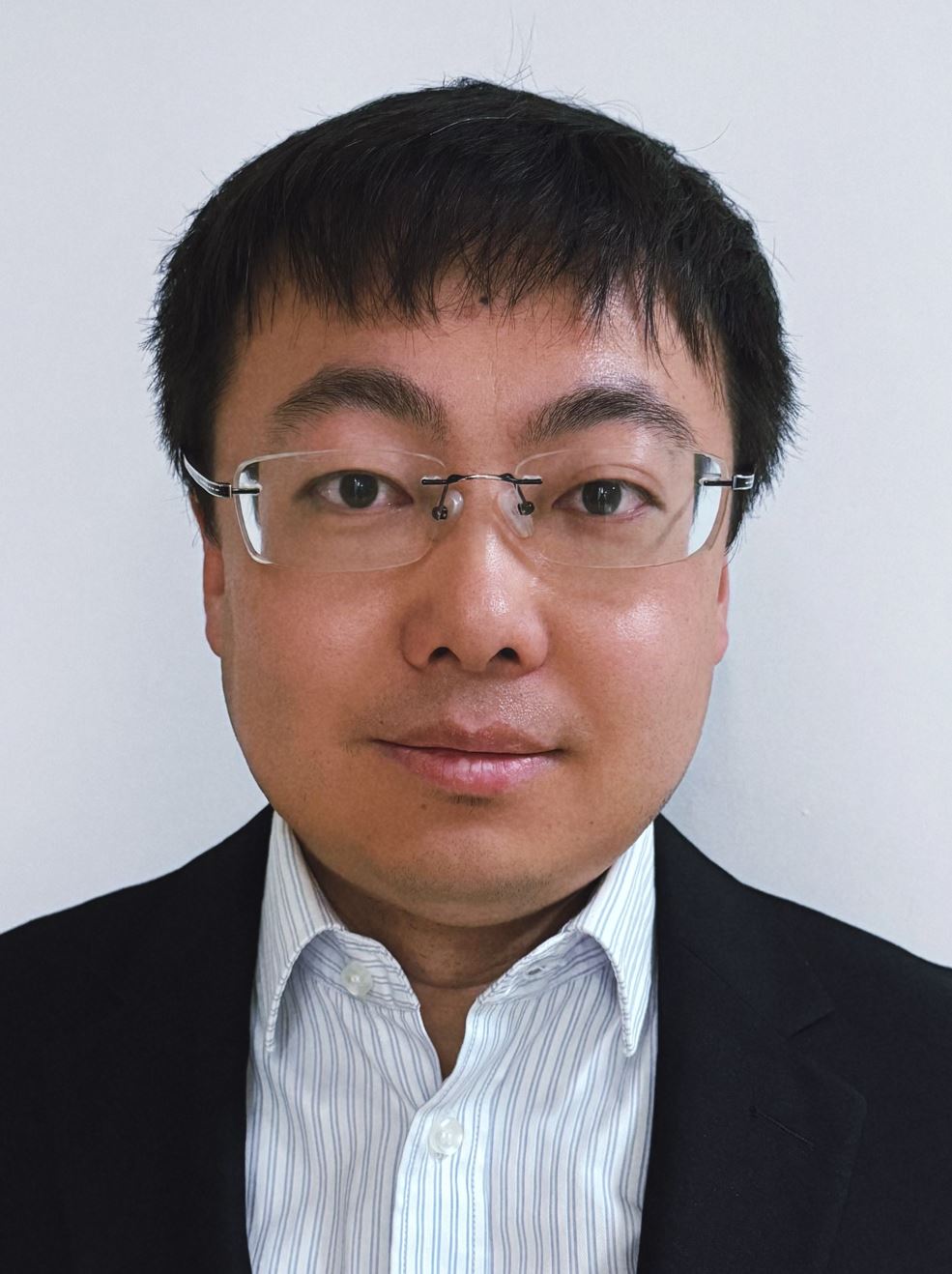}}]{Dequan Yu}
obtained his PhD from Dalian University of Technology in 2016. During his doctoral studies, he conducted research on the theoretical aspects of small molecule reaction dynamics. He is currently employed at Huawei Communication Technology Co., Ltd. as an algorithm engineer, where he is engaged in the development of deep learning image and video processing techniques. Since 2021, he has played an active role in the JPEG AI standardization process, which was issued by ISO/IEC. 
\end{IEEEbiography}
\begin{IEEEbiography}[{\includegraphics[width=1in,height=1.25in,clip,keepaspectratio]{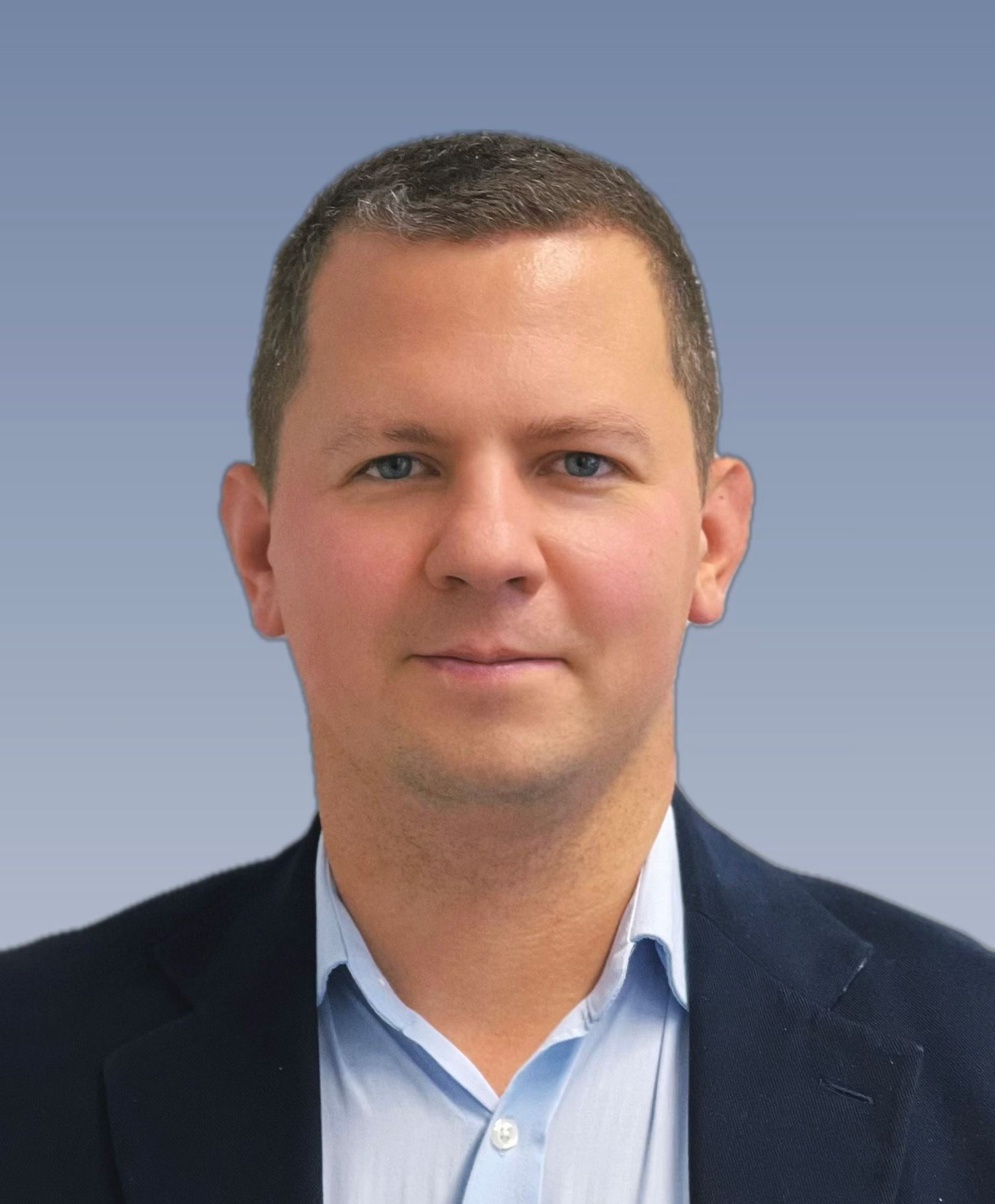}}]{Alexander Karabutov}
graduated from Moscow State University and got his PhD in Physics in 2013. In 2016 he joined Huawei Technologies in Moscow Research Center. Participated in Huawei's responce on VVC/H.266 Call for Proposals and EVC/MPEG-5. Since 2020 joined JPEG-AI standardization activity the first standard of image compression by neural networks. He became the editor of one of the part of the related stadard. Since 2022 he is Principal Engineer in Munich Research Center of Huawei Technologies. He has authored and co-authored of papers in various journals and international conferences, 50+ patents. His current research interests include AI-based image and video compression.
\end{IEEEbiography}
\begin{IEEEbiography}[{\includegraphics[width=1in,height=1.25in,clip,keepaspectratio]{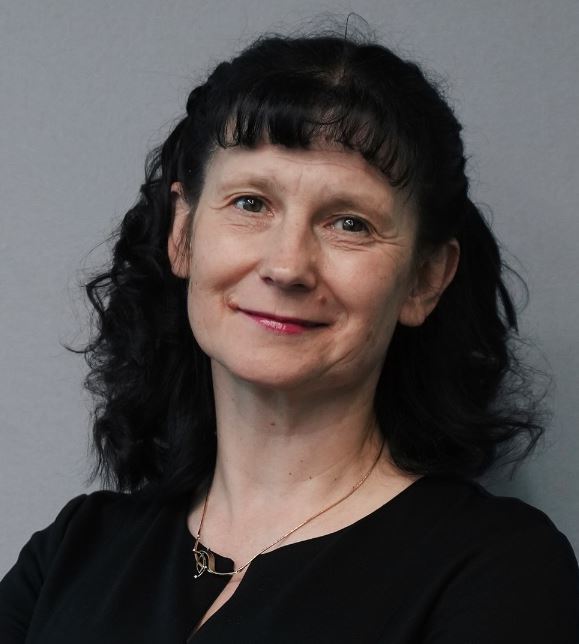}}]{Elena Alshina}
graduated from Moscow State University, got her PhD in Mathematical Modelling in 1998. She (together with Alexander Alshin) got the Gold medal of the Russian Academy of Science in 2003 for series of research papers on Computer Science. In 2006 she joined Samsung Electronics and was part of the team that submitted the topperforming response for HEVC/H.265 Call for Proposals in 2010. Since that time she is an active participant in international video codec standardization. She was chairing multiple core experiments and AhGs in JCTVC, JVET, MPEG. Since 2018 she is Chief Video Scientist in Huawei Technologies (Munich), also the director for Media Codec and Audiovisual technology Labs. She has authored several scientific books, more than 100 papers in various journals and international conferences, 200+ patents. She currently is a co-chair of JVET Exploration on Neural Network-based video coding and JPEG AI standardization project co-chair and co-editor. Her current research interests include AI-based image and video coding, signal processing, computer vision.
\end{IEEEbiography}
\begin{IEEEbiography}[{\includegraphics[width=1in,height=1.25in,clip,keepaspectratio]{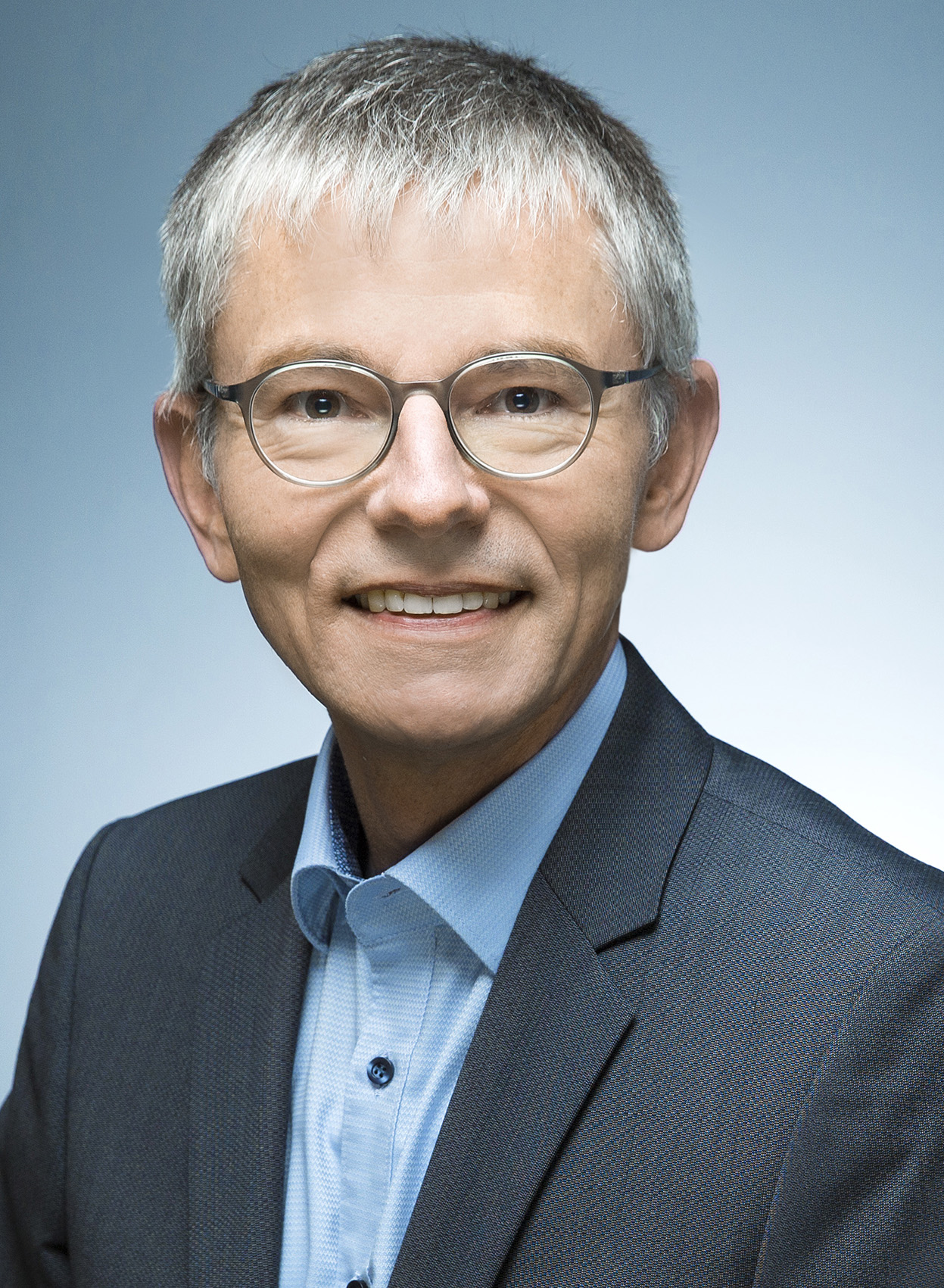}}]{André Kaup}
(Fellow, IEEE) received the Dipl.-Ing. and Dr.-Ing. degrees in electrical engineering from RWTH Aachen University, Aachen, Germany, in 1989 and 1995, respectively.
He joined Siemens Corporate Technology, Munich, Germany, in 1995, and became the Head of the Mobile Applications and Services Group in 1999. Since 2001, he has been a Full Professor and the Head of the Chair of Multimedia Communications and Signal Processing at Friedrich-Alexander University Erlangen-Nürnberg (FAU), Germany. From 2005 to 2007 he was Vice Speaker of the DFG Collaborative Research Center 603. From 2015 to 2017, he served as the Head of the Department of Electrical Engineering and Vice Dean of the Faculty of Engineering at FAU. He has authored around 500 journal and conference papers and has over 120 patents granted or pending. His research interests include image and video signal processing and coding, and multimedia communication.
Dr. Kaup is a member of the IEEE Image, Video, and Multidimensional Signal Processing Technical Committee and a member of the Scientific Advisory Board of the German VDE/ITG. He is an IEEE Fellow and a member of the Bavarian Academy of Sciences and Humanities, the German National Academy of Science and Engineering, and the European Academy of Sciences and Arts. He is a member of the Editorial Board of the IEEE Circuits and Systems Magazine. He was a Siemens Inventor of the Year 1998 and obtained the 1999 ITG Award. He received several IEEE best paper awards, including the Paul Dan Cristea Special Award in 2013, and his group won the Grand Video Compression Challenge from the Picture Coding Symposium 2013. The Faculty of Engineering with FAU and the State of Bavaria honored him with Teaching Awards, in 2015 and 2020, respectively. He served as an Associate Editor of the IEEE Transactions on Circuits and Systems for Video Technology. He was a Guest Editor of the IEEE Journal of Selected Topics in Signal Processing. 
\end{IEEEbiography}

\end{document}